\documentclass[a4paper,USenglish,cleveref, autoref, thm-restate]{lipics-v2021}

\hideLIPIcs  

\title{Linear Objective Problems: A Complexity Framework for Bilevel Optimization}

\author{Christoph Grüne}{Faculty of Computer Science, RWTH Aachen University, Germany}{gruene@algo.rwth-aachen.de}{https://orcid.org/0000-0002-7789-8870}{}
\author{James B. Orlin}{MIT Sloan School of Management, Cambridge, MA}{jorlin@mit.edu}{https://orcid.org/0000-0002-7488-094X}{}
\author{Lasse Wulf}{Department of Mathematics and Computer Science, University of Southern Denmark, Denmark}{lwulf@imada.sdu.dk}{https://orcid.org/0000-0001-7139-4092}{}

\authorrunning{C. Grüne, J.B. Orlin, L. Wulf}

\Copyright{Christoph Grüne, James B. Orlin, and Lasse Wulf} 



\category{}

\relatedversion{A preliminary version of parts of this work appeared at IPCO 2025~\cite{DBLP:conf/ipco/GruneW25}.}

\acknowledgements{}

\nolinenumbers

\EventEditors{John Q. Open and Joan R. Access}
\EventNoEds{2}
\EventLongTitle{42nd Conference on Very Important Topics (CVIT 2016)}
\EventShortTitle{CVIT 2016}
\EventAcronym{CVIT}
\EventYear{2016}
\EventDate{December 24--27, 2016}
\EventLocation{Little Whinging, United Kingdom}
\EventLogo{}
\SeriesVolume{42}
\ArticleNo{23}

\usepackage{amsmath,amsfonts,amssymb}
\usepackage{color}
\usepackage{hyperref}
\usepackage[T1]{fontenc}
\usepackage{graphicx}
\usepackage{csquotes}

\usepackage{tikz}
\usetikzlibrary{calc}

\usepackage{cleveref}
\crefname{claim}{claim}{claims}
\Crefname{claim}{Claim}{Claims}

\DeclareMathOperator{\reg}{reg}

\DeclareMathOperator*{\argmin}{arg\,min}

\newcommand{\set}[1]{\left\{#1\right\}}

\makeatletter
\newcommand{\comitem}[2]{%
  \item[#1]%
  \MakeLinkTarget[item]{}%
  \protected@edef\@currentlabel{#1}%
  \label{#2}%
}
\makeatother

\newcommand{\subjectto}[0]{\text{s.t.}}

\newcommand{\stconstraint}[4]{\subjectto \quad && #1 \quad & #2 \quad && #3 \qquad &&
\ifthenelse{\equal{#4}{}}{}{(\begin{array}{l} #4 \end{array})}}
\newcommand{\constraint}[4]{&& #1 \quad & #2 \quad && #3 \qquad &&
\ifthenelse{\equal{#4}{}}{}{(\begin{array}{l} #4 \end{array})}}

\newcommand{\R}{\mathbb{R}}

\newcommand{\Z}{\mathbb{Z}}

\newcommand{\powerset}[1]{2^{#1}}

\newcommand{\fromto}[2]{\set{#1, \ldots, #2}}
\newcommand{\bin}[0]{\set{0,1}}

\newcommand{\PTIME}{\textsf{P}}
\newcommand{\NP}{\textsf{NP}}
\newcommand{\coNP}{\textsf{coNP}}

\newcommand{\NPS}{\textsf{NP-S}}
\newcommand{\LOP}{\textsf{LOP}}
\newcommand{\CLOP}{\textsf{1-LOP}}

\newcommand{\leqSE}{\leq_{\mathrm{SE}}}

\newcommand{\leqTSE}{\leq_{\mathrm{TSE}}}

\newcommand{\U}{\mathcal{U}}

\newcommand{\sol}{\mathcal{S}}
\newcommand{\I}{\mathcal{I}}
\newcommand{\Uemb}{\U_{emb}}

\newif\ifnotes
\notestrue
\ifnotes
\newcommand{\todo}[1]{\textcolor{red}{\textbf{TODO: #1}}\\}
\newcommand{\important}[1]{\textcolor{purple}{\textsc{IMPORTANT: #1}}\\}

\newcommand{\notename}[2]{{\footnotesize{\bf (#1:} {#2}{\bf ) }}}
\newcommand{\jnote}[1]{\textcolor{red}{{\notename{Jim}{#1}}}}
\newcommand{\cnote}[1]{\textcolor{purple}{{\notename{Christoph}{#1}}}}
\newcommand{\lnote}[1]{{\textcolor{blue}{\notename{Lasse}{#1}}}}
\newcommand{\christoph}[1]{\textcolor{magenta}{C: #1 }}
\newcommand{\lasse}[1]{\textcolor{orange}{L: #1 }}
\newcommand{\jim}[1]{\textcolor{purple}{J: #1 }}
\newcommand{\citationneeded}[1][None]{\textsuperscript{\color{red} [Citation needed: #1]}}
\else
\newcommand{\todo}[1]{}
\newcommand{\important}[1]{}

\newcommand{\notename}[2]{}
\newcommand{\jnote}[1]{}
\newcommand{\cnote}[1]{}
\newcommand{\lnote}[1]{}
\newcommand{\christoph}[1]{}
\newcommand{\lasse}[1]{}
\newcommand{\jim}[1]{}
\newcommand{\citationneeded}[1][None]{}
\fi

\begin{document}

\maketitle
\begin{abstract}
    Many bilevel and robust optimization problems are complete for the second level of the polynomial hierarchy, but such results are usually proven one problem at a time with problem-specific reductions.
A recurring obstacle is that the lower level must select an \emph{optimal} solution rather than merely a feasible one, whereas standard reductions preserve only whether a solution exists, not which solutions are optimal.
We introduce \emph{linear objective problems} ($\LOP$), a framework for optimization problems whose solutions are explicit subsets of a ground set and whose objective is linear in these elements.
We define \emph{tight solution-embedding reductions}, which map the optimal solutions of one problem element by element onto the optimal solutions of another, without having to recognize optimal solutions, which is $\sf coNP$-hard in general.
This gives a notion of $\LOP$-completeness, with \textsc{Satisfiability} as the canonical complete problem, even with unit weights.

We show that many classical problems are $\LOP$-complete, including \textsc{Vertex Cover}, \textsc{Dominating Set}, \textsc{TSP}, \textsc{Knapsack}, and \textsc{Steiner Tree}, and we introduce a duality that transfers $\LOP$-completeness between minimization and maximization problems.
As our main application, we prove a meta-theorem for min-max regret optimization under interval uncertainty:
for every $\LOP$-complete problem, the min-max regret version is $\Sigma_2^p$-complete.
Instead of a separate proof for each nominal problem, it suffices to show that the nominal problem is $\LOP$-complete.
This recovers known results for the interval min-max regret versions of \textsc{Knapsack}~\cite{DBLP:journals/disopt/DeinekoW10} and \textsc{Maximum Benefit Set Cover}~\cite{DBLP:journals/coap/CocoSN22} and yields completeness for over 50 further problems studied in \cite{DBLP:journals/corr/abs-2511-04308}.

\end{abstract}

\section{Introduction}
\label{sec:introduction}

In recent years, there has been enormous interest in bilevel optimization \cite{dempe2020bilevel,DBLP:journals/ejco/KleinertLLS21},
robust optimization \cite{DBLP:books/degruyter/Ben-TalGN09,robook}, network interdiction \cite{smith2013modern}, Stackelberg games \cite{li2017review}, attacker-defender games \cite{HUNT2023}, and many other hierarchical optimization problems.
These problems arise whenever decisions are made sequentially or hierarchically, so that decisions at one level influence the choices
available to subsequent levels.
They provide natural models for situations in which a decision maker must anticipate the response of another decision maker, or in which decisions must remain effective under uncertain future conditions.

A central feature of these problems is that the computational structure of a level depends on how its response is determined.
In some multilevel problems, a level selects a feasible solution from a set determined by the decisions made at previous levels.
In others, the level must solve an optimization problem of its own, and its response is therefore required to be optimal with respect to its objective function.
This distinction is important not only for the mathematical formulation of a problem, but also for understanding its computational complexity.

A generic bilevel optimization problem can be written as
\begin{equation}
\label{eq:generic-bilevel}
    \min_{x \in X}
    \left\{
        w_1(x,y) :
        y \in \argmin_{y' \in Y(x)} w_2(x,y')
    \right\}.
\end{equation}
Here, $x$ denotes an upper-level decision, while $y$ denotes the response of the lower level.
The upper level chooses its decision while anticipating that the lower level will respond optimally to it.
Thus, the defining feature of \eqref{eq:generic-bilevel} is that the lower level does not merely select a feasible response.
Rather, it selects a response that minimizes its own objective function.

To understand the significance of this distinction, fix an upper-level decision $x$ and let $\mathcal{F}(x)$ denote the feasible responses of the lower level.
In a feasibility-based model, the relevant condition is simply
$$
    y \in \mathcal{F}(x).
$$
In an optimization-based model, the condition is instead
$$
    y \in \argmin_{y' \in \mathcal{F}(x)} w_2(x,y').
$$
The objective function therefore determines which feasible solutions are relevant as responses to the upper level.
A complexity theory that represents only the feasible solution set does not, by itself, capture this objective-induced distinction.

This distinction occurs naturally in a variety of hierarchical and adversarial optimization problems, including network interdiction and multistage adjustable robust optimization.
In such problems, subsequent decisions may depend on a choice made at an earlier level, and the structure of the possible responses is essential to the analysis.
When a level must additionally optimize an objective function, however, the feasible solution structure alone is insufficient:
the objective determines which of the feasible responses qualify as optimal.

A particularly transparent example is min-max regret optimization.
Let $\mathcal{F}$ be a feasible solution set, and let $c$ be a realization of uncertain costs.
The regret of a solution $x \in \mathcal{F}$ under $c$ is defined as
$$
    \reg(x,c)
    =
    c(x)-\min_{y\in\mathcal{F}}c(y).
$$
The second term is crucial.
It is the value of an optimal solution for the realized cost vector, rather than the value of an arbitrary feasible solution.
The min-max regret problem consequently has the nested structure
$$
    \min_{x\in\mathcal{F}}
    \max_{c\in C}
    \left(
        c(x)-\min_{y\in\mathcal{F}}c(y)
    \right),
$$
where $C$ denotes the set of possible cost realizations.
The inner minimization determines the benchmark against which the upper-level solution is evaluated.
Thus, even though the nominal optimization problem is defined by the feasible set $\mathcal{F}$, its role in the regret formulation depends on which solutions are optimal for each realization of the uncertain costs.
Min-max regret therefore illustrates a broader issue in multilevel optimization:
the complexity of a problem may depend not only on the structure of feasible solutions, but also on the structure induced by the objective function.

\subparagraph*{The computational challenge of the second quantifier.}
The computational challenges associated with such nested optimization problems have been emphasized by Woeginger \cite{DBLP:journals/4or/Woeginger21} in his article \emph{The Trouble with the Second Quantifier}.
The paper discusses the distinction between problems characterized by a single existential quantifier and those whose natural formulations involve an existential quantifier followed by a universal quantifier.
While the former are closely associated with the complexity class $\NP$, the latter naturally lead to the second level of the polynomial hierarchy, in particular to $\Sigma_2^p$.
Woeginger illustrates these difficulties through examples from combinatorial optimization and computational social choice.
One such example originates in social choice theory and concerns the computation of a bilinear program.
Kurz and Napel \cite{DBLP:journals/ol/KurzN16} posed the underlying problem as a challenge to the operations research community. Despite the small size of its input, the problem has proved remarkably difficult.
Woeginger uses this example to illustrate the gap between the success of established methods for $\NP$-complete problems and the challenges posed by problems involving a second quantifier.
The same concern is relevant to bilevel and robust optimization.
Over several decades, the operations research community has developed sophisticated integer programming techniques for solving many $\NP$-hard problems.
However, these techniques do not automatically extend to problems at the second level of the polynomial hierarchy.
In particular, the presence of a second quantifier can fundamentally change the complexity of a problem and the kinds of techniques that are needed to analyze it.

Similar observations motivated a Dagstuhl seminar on optimization at the second level of the polynomial hierarchy.
Its report \cite{DBLP:journals/dagstuhl-reports/BrotcorneBHH22} emphasizes that methodologies developed for $\NP$-complete problems do not
directly apply to robust and bilevel optimization problems, and that new techniques and theoretical insights are needed to understand this area.
These observations motivate the development of systematic complexity-theoretic tools for bilevel and multilevel optimization.
Min-max regret provides a natural setting in which to investigate this question, since its complexity has been studied for a broad range of nominal optimization problems.

\subparagraph*{Complexity of min-max regret.}
The min-max regret criterion is widely used in robust optimization and has been investigated for many classical optimization problems.
For nominal problems that can be solved in polynomial time, the complexity of their min-max regret counterparts has been studied for 
shortest path and minimum spanning tree \cite{DBLP:journals/dam/AverbakhL04},
minimum cut and $s$-$t$ cut \cite{DBLP:journals/disopt/AissiBV08},
1-center and 1-median problems with uncertainty in node or edge weights \cite{AVERBAKH1997247,DBLP:journals/dam/Averbakh03,DBLP:journals/informs/AverbakhB00},
and various scheduling problems \cite{lebedev2006complexity}.
For interval uncertainty,
minimum cut,
1-center with uncertainties on nodes,
and 1-median with uncertainties on nodes
admit polynomial-time algorithms, whereas the others are shown to be $\NP$-hard.

Min-max regret has also been investigated for nominal problems that are themselves $\NP$-complete.
Relevant examples include knapsack \cite{DBLP:journals/disopt/DeinekoW10}, set cover \cite{DBLP:journals/coap/CocoSN22}, and the traveling salesman and Steiner tree problems \cite{DBLP:journals/talg/GaneshMP23}.
More general treatments of min-max regret and robust optimization can be found in \cite{kouvelis2013robust,kasperski2016robust}, and a comprehensive survey is provided by Aissi, Bazgan, and Vanderpooten \cite{DBLP:journals/eor/AissiBV09}.
Of particular relevance to the present work are completeness results at the second level of the polynomial hierarchy.
Deineko and Woeginger \cite{DBLP:journals/disopt/DeinekoW10} established $\Sigma_2^p$-completeness for interval min-max regret knapsack, while Coco, Santos, and Noronha \cite{DBLP:journals/coap/CocoSN22} established $\Sigma_2^p$-completeness for interval min-max regret maximum benefit set cover.
These results show that min-max regret can introduce a substantial increase in computational complexity and provide concrete examples of second-level completeness in robust optimization.

These completeness results each concerned a min-max regret variant of an $\NP$-complete problem.  Yet each required problem-specific arguments, even when the resulting regret problems exhibited similar complexity behavior.  It led to the following natural question:
\emph{Can one develop a methodology or framework for establishing completeness results for bilevel optimization problems in which a lower level must select an optimal solution rather than merely a feasible one?}  

For min-max regret, a more specific version of this question is whether the completeness of a nominal optimization problem can be transferred systematically to its min-max regret counterpart.
Answering this question requires a framework that preserves not only the feasible solutions of a nominal problem, but also the objective-induced structure that determines its optimal solutions.

\subparagraph*{Solution-based complexity frameworks.}
A recent line of work develops systematic approaches to completeness proofs for problems at higher levels of the polynomial hierarchy.
Johannes \cite{Johannes11_CompleteProblems2LevelPolyHierarchy} developed techniques for obtaining families of $\Sigma_2^p$-complete
problems.
More recently, in a preliminary version of this work \cite{DBLP:conf/ipco/GruneW25}, Grüne and Wulf introduced meta-theorems establishing $\Sigma_2^p$- and $\Sigma_3^p$-completeness for broad classes of bilevel and robust optimization problems.
Related techniques and complexity results have also been developed for specific classes of bilevel and multilevel optimization problems, including Stackelberg pricing games \cite{DBLP:conf/esa/GruneHRW26}, blocking and interdiction problems \cite{DBLP:journals/tcs/GruneW26}, and recoverable robust optimization \cite{DBLP:conf/mfcs/GruneW25}.
Subsequently, Grüne, Johannes, Orlin, and Wulf \cite{DBLP:journals/corr/abs-2602-12350} introduced the complexity class \emph{$\NP$ with solutions}, denoted by $\NPS$, providing a formal framework for capturing the solution structure underlying such problems.
In contrast to the usual language-based perspective of $\NP$, the $\NPS$ framework explicitly represents a universe of solution
elements and a set of solutions.
This representation allows reductions to preserve not only the existence of solutions, but also the correspondence between solution
elements of the source and target problems.
Such reductions are called \emph{solution-embedding reductions}.
A compendium of reductions \cite{DBLP:journals/corr/abs-2511-04308} collects problems and solution-embedding reductions as well as other reductions related to the above papers, providing a common reference for techniques relevant to these developments.

The $\NPS$ framework provides a natural starting point for analyzing the complexity of multilevel optimization.
However, it does not directly distinguish between feasible solutions according to their objective values.
Given a feasible solution set $\mathcal{F}$ and an objective function $f$, an $\NPS$ representation records the solutions in $\mathcal{F}$ but does not, by itself, identify the subset
$$
    \arg\min_{y\in\mathcal{F}} f(y)
$$
as being structurally different from the remaining feasible solutions.
Consequently, a solution-embedding reduction may preserve the complete feasible solution structure while changing the objective values of the solutions and, therefore, changing which solutions are optimal.
For feasibility-based multilevel problems, preserving the feasible solution structure is the relevant requirement.
For optimization-based multilevel problems, however, this is not sufficient:
the response of a lower level is determined by optimality, and this distinction must be preserved by the reductions used in a completeness proof.

\paragraph*{Contributions}

In this paper, we extend the solution-based perspective to optimization problems by introducing \emph{linear objective problems}.
A linear objective problem consists of an $\NPS$ problem together with a linear objective function over its universe elements.
The objective function is an essential part of the problem structure because it determines which feasible solutions are optimal and, consequently, which solutions may serve as responses at a subsequent level.

The purpose of linear objective problems is not merely to attach weights to $\NPS$ problems, but to provide a framework in which the objective-induced structure of optimal solutions can be preserved by reductions.
We denote the corresponding complexity class by $\LOP$ and introduce \emph{tight solution-embedding reductions}. 
These reductions extend the solution-embedding property with a tightness condition that preserves objective thresholds identifying optimal solutions.
This allows reductions to preserve the structure between optimal solutions without requiring them to recognize or compute optimal solutions, which may itself be computationally difficult.
Based on these reductions, we define \emph{$\LOP$-completeness}.
The crucial difference from $\NPS$-completeness is that $\LOP$-completeness preserves the objective-induced distinction between feasible and optimal solutions.
Thus, $\NPS$ is appropriate when a level is characterized by the selection of a feasible solution, whereas $\LOP$ is appropriate when a level is characterized by the solution of an optimization problem.

We then apply our framework to establish a general completeness result for min-max regret optimization under interval uncertainty.
For a realization of the uncertain data, the benchmark in the regret expression is an optimal solution of the nominal optimization problem.
The lower level is therefore a linear objective problem of the type captured by $\LOP$.
Our main application result is the following:
\begin{quote}
    For every $\LOP$-complete nominal optimization problem, the
    corresponding min-max regret problem with interval uncertainty
    is $\Sigma_2^p$-complete.
\end{quote}
This result provides a uniform transfer principle:
rather than requiring a separate completeness argument for each nominal problem, it suffices to establish that the nominal problem is $\LOP$-complete.
In particular, the result recovers known completeness results for specific min-max regret problems as special cases, including min-max regret knapsack \cite{DBLP:journals/disopt/DeinekoW10}
and min-max regret maximum benefit set cover \cite{DBLP:journals/coap/CocoSN22}.
The $\LOP$ framework makes this transfer possible because tight solution-embedding reductions preserve the optimal-solution structure of the nominal problem that determines the regret benchmark.
This transfer would not be possible if one relied only on the $\NPS$ framework.

Completeness in $\Sigma_2^p$ directly implies that standard integer linear programming approaches are insufficient for solving the problem.
If a $\Sigma_2^p$-complete problem admitted a polynomial-size integer linear programming formulation that can be computed in polynomial time, it would imply that the problem was in $\NP$.  In turn, this would imply that $\Sigma_2^p = \NP$ and hence a collapse of the polynomial hierarchy.
Thus, under standard complexity assumptions, none of the min-max regret problems covered by our result admits a compact integer linear programming formulation.
This explains why exact methods for min-max regret rely on formulations of exponential size and shows that this is unavoidable for every $\LOP$-complete nominal problem.

More broadly, $\LOP$ provides a foundation for completeness results that account explicitly for the role of optimality in hierarchical decision-making.
Moreover, this framework is not tied to min-max regret:
the completeness results for interdiction \cite{DBLP:journals/tcs/GruneW26}, Stackelberg pricing \cite{DBLP:conf/esa/GruneHRW26}, and recoverable robust optimization \cite{DBLP:conf/mfcs/GruneW25} rely on the same principle.

\subparagraph*{Relation to the conference version.}
A preliminary version of this work appeared at IPCO 2025~\cite{DBLP:conf/ipco/GruneW25}.
Its results have since been developed into two full versions:
\cite{DBLP:journals/corr/abs-2602-12350} introduces the class $\NPS$ for problems in which a level selects a feasible solution, and the present paper introduces $\LOP$ for problems in which a level solves an optimization problem.
Compared to~\cite{DBLP:conf/ipco/GruneW25}, the present paper contains the formal class $\LOP$ together with its completeness notion, the duality between minimization and maximization problems, complete proofs, and the compendium of $\LOP$-complete problems.

\paragraph*{Paper Organization}
The remainder of the paper is organized as follows.
In \Cref{sec:preliminaries}, we introduce notation and preliminary definitions.
\Cref{sec:framework} formally introduces the $\LOP$ framework, including linear objective problems, tight thresholds, tight solution-embedding reductions, and $\LOP$-completeness.
It also establishes the relationship between $\NPS$ and $\LOP$ and discusses relevant subclasses.
In \Cref{sec:regret-def}, we define min-max regret problems based on $\LOP$ problems and show containment in $\Sigma^p_2$.
\Cref{sec:regret-sat} considers the min-max regret version with interval uncertainty of \textsc{Satisfiability} and shows its $\Sigma^p_2$-completeness.
In \Cref{sec:regret-min}, the meta-theorem that the min-max regret version with interval uncertainty for minimization $\LOP$ problems is $\Sigma^p_2$-complete is proven.
\Cref{sec:regret-max} introduces a duality notion for $\LOP$ problems, which relates minimization $\LOP$ problems with maximization $\LOP$ problems and thus proves the meta-theorem for maximization $\LOP$ problems.
\Cref{sec:conclusion} concludes our paper.
In \Cref{sec:compendium}, a compendium of $\LOP$-complete problems can be found.

\section{Preliminaries}
\label{sec:preliminaries}

\subparagraph*{Complexity.}
A \emph{language} is a set $L\subseteq \{0,1\}^*$.
For a language $L$, $w$ is a yes-instance if $w \in L$; otherwise it is a no-instance.
A language $L$ is contained in $\Sigma^p_2$ if and only if there exists some polynomial-time computable function $V$ (also called \emph{verifier}) such that, for all $w \in \set{0,1}^*$ and suitable $m_1,m_2 = |w|^{O(1)}$,
$$
    w \in L \ \Leftrightarrow \ \exists y_1 \in \set{0,1}^{m_1} \ \forall y_2 \in \set{0,1}^{m_2}: V(w,y_1,y_2) = 1.
$$
An introduction to the polynomial hierarchy and the class $\Sigma^p_2$ can be found in the book by Papadimitriou \cite{DBLP:books/daglib/0072413} or in the article by Jeroslow \cite{DBLP:journals/mp/Jeroslow85}.
An introduction specifically in the context of bilevel optimization can be found in the article of Woeginger \cite{DBLP:journals/4or/Woeginger21}.

A \emph{polynomial-time many-one reduction} (or \emph{Karp reduction}) from a language $L$ to a language~$L'$ is a polynomial-time computable map $g : \{0,1\}^* \to \{0,1\}^*$ such that $w \in L$ if and only if $g(w) \in L'$ for all $w \in \{0,1\}^*$.
A language $L$ is $\Sigma^p_2$-hard if every $L' \in \Sigma^p_2$ can be reduced to $L$ with a polynomial-time many-one reduction. If $L$ is both $\Sigma^p_2$-hard and contained in $\Sigma^p_2$, it is $\Sigma^p_2$-complete.

\subparagraph*{Boolean formulas.}
A \emph{Boolean variable} $x$ is a variable that takes one of the values 0 or 1. Let $X = \fromto{x_1}{x_n}$ be a set of variables. The corresponding \emph{literal set} is given by $L = \fromto{x_1}{x_n} \cup \fromto{\overline x_1}{\overline x_n}$.
A \emph{clause} is a disjunction of literals. 
A Boolean formula is in \emph{conjunctive normal form} (CNF) if it is a conjunction of clauses.
It is in \emph{disjunctive normal form} (DNF) if it is a disjunction of \emph{product terms}, that is, of conjunctions of literals.
We use the convention that a clause or a product term may be represented by a subset of the literals, and that a CNF formula (DNF formula) $\varphi$ may be represented by a set of clauses (product terms). 
For example, the set $\set{\set{x_1, \overline x_2},\set{x_2, x_3}}$ corresponds to the CNF formula $(x_1 \lor \overline x_2)\land (x_2 \lor x_3)$.
We write $\varphi(X)$ to indicate that formula $\varphi$ depends only on $X$.
Sometimes we are interested in cases where the variables are partitioned. We indicate this case by writing $\varphi(X,Y,\ldots)$.

An \emph{assignment} of variables is a map $\alpha : X \to \bin$.
The evaluation of the formula $\varphi$ under assignment $\alpha$ is denoted by $\varphi(\alpha) \in \bin$.
We also denote a partial assignment $\alpha$ on $X$ of a formula with partitioned variables $\varphi(X, Y)$ with $\varphi(\alpha, Y)$.
We write $\alpha \vDash \varphi$ if $\alpha$ satisfies formula $\varphi$; that is, $\varphi(\alpha)=1$.
The \emph{complementary assignment} of $\alpha$ is $\overline\alpha := 1 - \alpha$.

Assignments correspond to sets of literals. For an assignment $\alpha$, we define
$$
    L_\alpha := \{x_i : \alpha(x_i) = 1\} \cup \{\overline x_i : \alpha(x_i) = 0\}.
$$
Note that $L \setminus L_\alpha = L_{\overline\alpha}$.
A clause $c \subseteq L$ is satisfied by $\alpha$ if $L_\alpha \cap c \neq \emptyset$, and a product term $p \subseteq L$ is satisfied by $\alpha$ if $p \subseteq L_\alpha$.
Accordingly, we define the \emph{solution set} of a CNF formula $\varphi$ with clause set $C$ by
$$
    \sol(\varphi) := \{S \subseteq L : |S \cap \{x_i, \overline x_i\}| = 1 \ \forall i \in \{1,\dots,n\},\ |S \cap c| \geq 1 \ \forall c \in C\},
$$
and the solution set of a DNF formula $\psi$ with product terms $T = \{p_1, \ldots, p_m\}$, where $p_j \subseteq L$ for all $j$, by
$$
    \sol(\psi) := \{S \subseteq L : |S \cap \{x_i, \overline x_i\}| = 1 \ \forall i \in \{1,\dots,n\},\ \exists j \in \{1,\dots,m\}: p_j \subseteq S\}.
$$
In both cases, $S \in \sol(\varphi)$ if and only if $S = L_\alpha$ for some assignment $\alpha$ with $\alpha \vDash \varphi$.

\subparagraph*{Sets and functions.}
For a finite set $U$, we denote by $\powerset{U}$ its power set.
A \emph{cost function} on a set $A$ is a mapping  $c : A \to \R$.
For each subset $A' \subseteq A$, we define the cost of the subset $A'$ as $c(A') := \sum_{a \in A'} c(a)$.
For a map $f : A \to B$ and some subset $A' \subseteq A$, we define the image of the subset $A'$ as $f(A') = \set{f(a) : a \in A'}$. For some $B' \subseteq B$, we denote its preimage by $f^{-1}(B') = \set{a \in A : f(a) \in B'}$.
In this paper, it will be apparent whether a function is a cost function or not, and so the meaning of the notation will be clear from context.
For a cost function $c$ and a family $\mathcal F \subseteq \powerset{A}$, we use the convention that $\arg\min_{S \in \mathcal F} c(S)$ and $\arg\max_{S \in \mathcal F} c(S)$ denote the \emph{sets} of all minimizers and maximizers, respectively.

\section{Framework}
\label{sec:framework}
In this section, we describe optimization problems in a way that allows us to formulate general reductions between them.
For this purpose, we consider alternative formulations of classical computational problems from $\NP$ in which solutions are made explicit.
We capture these problems in the complexity class \emph{$\NP$ with solutions} ($\NPS$), which enriches $\NP$ by semantically associating instances with explicit solution structures.

We then extend this framework to \emph{linear objective problems} ($\LOP$).
An $\LOP$ problem consists of an $\NPS$ problem together with a linear objective over its universe elements.
We distinguish between the optimization version, which selects feasible solutions according to their objective value, and the corresponding decision version, which asks whether there is a feasible solution whose objective value
satisfies a given threshold.

Our main goal is to use this framework to reason about optimal solutions.
In particular, we introduce \emph{tight solution-embedding reductions}, which strengthen solution-embedding reductions by additionally preserving the solution structure between optimal solutions.
Based on these reductions, we define a completeness notion for the complexity class $\LOP$.
The key property of this notion is that the optimal solutions of every problem in $\LOP$ can be mapped, element by element, onto the optimal solutions of any $\LOP$-complete problem, even though recognizing an optimal solution is in general $\coNP$-hard.

\subsection{NP with Solutions}
\label{sec:np-s}
We start by explaining the important concepts of the class $\NPS$, which was introduced by Grüne, Johannes, Orlin, and Wulf in \cite{DBLP:journals/corr/abs-2602-12350}.
In the standard definition of $\NP$, a problem is represented as a language, that is, a set of instances for which the answer to the decision question is yes.
A deterministic polynomial-time verifier checks whether a proposed polynomial-size certificate (or witness) proves that an instance is a yes-instance.
Once the verifier accepts, the complexity-theoretic definition of the language does not distinguish between different certificates that may have been used.
In particular, $\NP$ is concerned only with whether at least one valid certificate exists, and not with the structure or description of possible certificates.

\subparagraph*{Motivation for $\NPS$.}
Many combinatorial problems naturally have explicit solutions.
The class $\NPS$ changes this perspective by making the structure of these solutions part of the mathematical description of a problem.
Instead of representing solutions merely as arbitrary binary strings, a problem from $\NPS$ explicitly associates every instance with a universe of possible solution elements and with a set of solutions, where every solution is represented as a subset of that universe.
The decision question associated with an $\NPS$ problem is whether the solution set contains at least one feasible solution.
Thus, $\NPS$ does not change the underlying yes/no question of an $\NP$ problem.
Instead, it enriches its semantic description by retaining information about the individual solution elements and about which subsets of these elements constitute solutions.

The following is the formal definition.

\begin{definition}[Non-deterministic Polynomial Time with Solutions, from \cite{DBLP:journals/corr/abs-2602-12350}]
\label{def:NP-S}
A problem $\Pi = (\I, \U, \sol)$ is in $\NPS$ if there is a deterministic polynomial time Turing machine $V$, such that
\begin{itemize}
    \item $\I \subseteq \set{0,1}^*$ is a language. We call $\I$ the set of instances of $\Pi$. 
    \item To each instance $I \in \I$, there is some set $\U(I)$, such that $|\U(I)| = |I|^{O(1)}$, where $|I|$ denotes the length of the input string.  We refer to $\U(I)$ as the universe associated to the instance $I$. 
    \item To each instance $I \in \I$, there is some (potentially empty) set $\sol(I) \subseteq \powerset{\U(I)}$ that we call the solution set associated to the instance $I$.
    \item For each $C \subseteq \U(I)$, $V(I, C) = 1$ if and only if $C \in \sol(I)$.
\end{itemize}
\end{definition}
The decision question associated with an $\NPS$ problem is therefore:
Is $\sol(I) \neq \emptyset?$

As an example, let us consider \textsc{Satisfiability}, which we can express as an $\NPS$ problem as follows.
\medskip
\begin{description}
    \item[] \textsc{Satisfiability (Literal Universe)} as $\NPS$ problem \hfill\\
    \textbf{Instances $\I$:} \quad Literals $L = \fromto{\ell_1, \overline \ell_1}{\ell_n, \overline \ell_n}$, Clauses $C \subseteq 2^L$.\\
    \textbf{Universe $\U$:} \quad Literal set $L = \fromto{\ell_1, \overline \ell_1}{\ell_n, \overline \ell_n}$.\\
    \textbf{Solutions $\sol$:} \quad $S \subseteq L$ such that $|S \cap \{\ell_i,\overline \ell_i\}| = 1$ for all $i$ and $|c \cap S| \geq 1$ for all $c \in C$.
\end{description}

$\NPS$ does not enlarge the computational power of $\NP$;
it enriches the semantic representation of an $\NP$ problem.
The solution set is not explicitly enumerated as part of the input.
Instead, it is implicitly defined by the polynomial-time verifier.
However, the membership of a solution candidate in the solution set is polynomial-time decidable.
The distinction from ordinary $\NP$ is that $\NPS$ retains the interpretation of a certificate as a subset of a specified universe and hence preserves information about the individual elements occurring in a solution.

This distinction can also be formalized by relating $\NPS$ problems back to ordinary $\NP$ languages.

\begin{lemma}[from \cite{DBLP:journals/corr/abs-2602-12350}]
\label{lem:subsetNP}
    Each problem $\Pi$ in $\NPS$ can be transformed into a language $L_\Pi$ in $\NP$.
\end{lemma}

Intuitively, the construction incorporates a description of the verifier into the encoding of the instance.
A universal Turing machine can then simulate the corresponding verifier on a proposed solution, yielding an ordinary $\NP$ language.
Thus, $\NPS$ should be viewed as a richer representation of $\NP$ problems rather than as a larger complexity class.

\subparagraph*{The universe and solution set matter.}

Although $\NPS$ does not increase computational power, the additional structure leads to some important distinctions.  In particular, two $\NPS$ problems may represent the same underlying language of yes-instances of the original $\NP$ problem while being different $\NPS$ problems.
In ordinary $\NP$, two formulations that accept exactly the same instances represent the same language.
In $\NPS$, this is no longer sufficient: the universe of solution elements and the solution set are part of the problem's definition.

Consider \textsc{Clique} as a prototypical example.
In the natural \textsc{Clique (Vertex Universe)} formulation, the universe consists of the vertices of the input graph.
A solution is a subset of vertices that forms a clique.
Alternatively, we may consider \textsc{Clique (Edge Universe)}, in which the universe consists of the edges of the input graph.
A solution is then the set of edges induced by a clique.
Both formulations represent the same underlying $\NP$ decision problem:
whether the graph contains a clique of the required size.
However, they are different $\NPS$ problems because their universes, and consequently their solution objects, are different.

A similar phenomenon occurs for \textsc{Satisfiability}.
One may define \textsc{Satisfiability (Literal Universe)}, abbreviated \textsc{Sat-L}, in which the universe consists of literals, or \textsc{Satisfiability (Variable Universe)}, abbreviated \textsc{Sat-V}, in which the universe consists of variables and a solution is the set of variables set to~$1$.
Again, these formulations can encode the same underlying $\NP$ decision problem while giving different meanings to the elements of a solution.

The solution set itself is part of the relevant structure.
Consider the problem \textsc{Partition}.
Given a set of natural numbers $A = \{a_1, \ldots, a_n\}$,
a solution is a subset $A' \subseteq A$ such that $\sum_{a \in A'} a = \sum_{a \notin A'} a$.
If every such subset is considered a solution, then the complement of each solution is another solution.
We could instead define a different $\NPS$ formulation in which only solutions containing $a_n$ are allowed.
This does not change the underlying $\NP$ problem:
whenever a partition exists, exactly one of the two complementary parts contains $a_n$.
Nevertheless, it changes the solution set and therefore gives a different $\NPS$ problem.

These examples illustrate a central feature of $\NPS$:
the class distinguishes between different representations of the same $\NP$ language whenever they differ in the structure of their solutions.

\subparagraph*{Solution-embedding reductions.}
We will later rely on polynomial-time many-one reductions that preserve the structure of individual solutions.
The usual many-one reductions preserve only the existence of solutions, and not the structure.
They do not preserve how individual solution elements correspond between the source and target problem.
Therefore, \cite{DBLP:journals/corr/abs-2602-12350} introduces a more restrictive notion of reduction called \emph{solution-embedding (SE) reduction}.
An SE reduction embeds the universe of the source problem into the universe of the target problem such that the universe elements in the embedded part of the target solutions correspond one-to-one to universe elements of the source problem.

We start by explaining the concept with an example.
For this, consider the $\NPS$ problem \textsc{Satisfiability (Literal Universe)} from above and the problem Directed Hamiltonian Path, which is defined as follows:
\medskip
\begin{description}
    \item[]\textsc{Directed Hamiltonian Path} \hfill\\
    \textbf{Instances $\I$:} \quad Directed graph $G = (V, A)$, vertices $s, t \in V$.\\
    \textbf{Universe $\U$:} \quad Arc set $A$.\\
    \textbf{Solutions $\sol$:} \quad The set of all $S \subseteq A$ such that $S$ is an $s$-$t$-path of length $|V|-1$.
\end{description}

Arora and Barak \cite{DBLP:books/daglib/0023084} present a reduction from \textsc{Satisfiability} to \textsc{Directed Hamiltonian Path}, which we describe in the following.
An example is depicted in \Cref{fig:reduction:3sat-directed-hamiltonian-path}.
First, they introduce a designated start vertex $s$ and a designated terminal vertex $t$.
Then, for each complementary literal pair $(\ell, \overline \ell)$, they introduce a long chain of vertices with distinct left and right endpoints that can be traversed in both directions.
All of the chains are linked together sequentially between $s$ and $t$.
Finally, they introduce the clause vertices.
Each clause is represented by a single vertex and is connected to the chain as follows:
If a positive literal is in a clause, they add arcs to and from its chain that can only be used if the chain is traversed from left to right.
If a negative literal is in a clause, they add arcs to and from the same chain that can only be used if the chain is traversed from right to left.

Arora and Barak then prove that a solution $S$ of the \textsc{Satisfiability} instance can be transformed into a Hamiltonian path from $s$ to $t$. This path sequentially traverses the literal chains from left to right if $S$ contains the positive literal $\ell$, and from right to left if $S$ contains the negative literal $\overline \ell$.
While traversing the chains, the path detours to visit each clause vertex exactly once via one of its satisfying literals.

\tikzstyle{vertex}=[draw,circle,fill=black, minimum size=4pt,inner sep=0pt]
\tikzstyle{edge} = [draw,-]
\tikzstyle{arc} = [draw,->]
\tikzstyle{doublearc} = [draw,<->]
\begin{figure}[thpb]
\centering
\begin{tikzpicture}[scale=0.6,auto]

\node[vertex] (s) at (2.5,5) {}; \node[above] at (s) {$s$};
\node[vertex] (t) at (2.5,1) {}; \node[below] at (t) {$t$};

\node[] (x1s) at (-2.5,4) {}; \node[right] at (x1s) {$\ell_1, \overline \ell_1$};
\node[vertex] (x11) at ($(x1s) + (2,0)$) {};
\node[vertex] (x12) at ($(x1s) + (4,0)$) {};
\node[vertex] (x13) at ($(x1s) + (6,0)$) {};
\node[vertex] (x14) at ($(x1s) + (8,0)$) {};
\draw[arc, bend left, looseness=0.5] (x11) to (x12);
\draw[arc, bend left, looseness=0.5] (x12) to (x13);
\draw[arc, bend left, looseness=0.5] (x13) to (x14);
\draw[arc, bend left, looseness=0.5] (x14) to (x13);
\draw[arc, bend left, looseness=0.5] (x13) to (x12);
\draw[arc, bend left, looseness=0.5] (x12) to (x11);

\node[] (x2s) at (-2.5,3) {}; \node[right] at (x2s) {$\ell_2, \overline \ell_2$};
\node[vertex] (x21) at ($(x2s) + (2,0)$) {};
\node[vertex] (x22) at ($(x2s) + (4,0)$) {};
\node[vertex] (x23) at ($(x2s) + (6,0)$) {};
\node[vertex] (x24) at ($(x2s) + (8,0)$) {};
\draw[arc, bend left, looseness=0.5] (x21) to (x22);
\draw[arc, bend left, looseness=0.5] (x22) to (x23);
\draw[arc, bend left, looseness=0.5] (x23) to (x24);
\draw[arc, bend left, looseness=0.5] (x24) to (x23);
\draw[arc, bend left, looseness=0.5] (x23) to (x22);
\draw[arc, bend left, looseness=0.5] (x22) to (x21);

\node[] (x3s) at (-2.5,2) {}; \node[right] at (x3s) {$\ell_3, \overline \ell_3$};
\node[vertex] (x31) at ($(x3s) + (2,0)$) {};
\node[vertex] (x32) at ($(x3s) + (4,0)$) {};
\node[vertex] (x33) at ($(x3s) + (6,0)$) {};
\node[vertex] (x34) at ($(x3s) + (8,0)$) {};
\draw[arc, bend left, looseness=0.5] (x31) to (x32);
\draw[arc, bend left, looseness=0.5] (x32) to (x33);
\draw[arc, bend left, looseness=0.5] (x33) to (x34);
\draw[arc, bend left, looseness=0.5] (x34) to (x33);
\draw[arc, bend left, looseness=0.5] (x33) to (x32);
\draw[arc, bend left, looseness=0.5] (x32) to (x31);

\draw[arc,in=90,out=225, looseness=0.25] (s) to (x11);
\draw[arc,in=90,out=315, looseness=0.25] (s) to (x14);
\draw[arc,in=90,out=270, looseness=0.25] (x11) to (x21);
\draw[arc,in=90,out=270, looseness=0.25] (x11) to (x24);
\draw[arc,in=90,out=270, looseness=0.25] (x14) to (x21);
\draw[arc,in=90,out=270, looseness=0.25] (x14) to (x24);
\draw[arc,in=90,out=270, looseness=0.25] (x21) to (x31);
\draw[arc,in=90,out=270, looseness=0.25] (x21) to (x34);
\draw[arc,in=90,out=270, looseness=0.25] (x24) to (x31);
\draw[arc,in=90,out=270, looseness=0.25] (x24) to (x34);
\draw[arc,in=180,out=270, looseness=0.25] (x31) to (t);
\draw[arc,in=0,out=270, looseness=0.25] (x34) to (t);

\node[vertex] (c1) at (10,3) {}; \node[right] at (c1) {$c_1 = \overline \ell_1 \lor \overline \ell_2 \lor \ell_3$};
\draw[arc, bend left, dashed, looseness=0.75] (x13) to (c1);
\draw[arc, bend right, dashed, looseness=0.75] (c1) to (x12);
\draw[arc, bend left, dashed, looseness=0.75] (x23) to (c1);
\draw[arc, bend right, dashed, looseness=0.75] (c1) to (x22);
\draw[arc, bend right, dashed, looseness=0.75] (x32) to (c1);
\draw[arc, bend left, dashed, looseness=0.75] (c1) to (x33);

\end{tikzpicture}
\caption{Classical reduction of \textsc{Satisfiability} to \textsc{Directed Hamiltonian Path} for $\varphi = (\overline \ell_1 \vee \overline \ell_2 \vee \ell_3)$.}
\label{fig:reduction:3sat-directed-hamiltonian-path}
\end{figure}
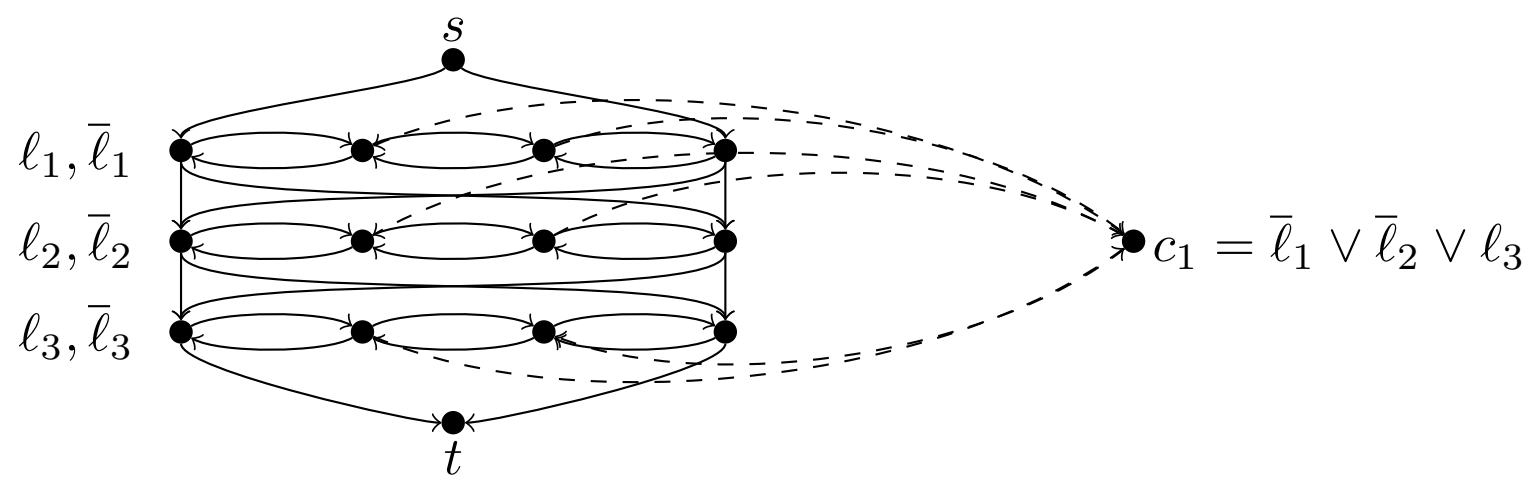

The important observation is that not only the existence of solutions is preserved, but the solution structure of the \textsc{Satisfiability} instance is mapped into the solution structure of the \textsc{Directed Hamiltonian Path} instance.
To make this precise, we designate two arcs for the chain of the pair $(\ell_i, \overline\ell_i)$: the arc $a_i^{\rightarrow}$ from the left endpoint to its neighbor, and the arc $a_i^{\leftarrow}$ from the right endpoint to its neighbor.
We attach the clause detours away from the endpoints of each chain, which can always be ensured by adding one buffer vertex at each end of the chain.
Then no detour bypasses $a_i^{\rightarrow}$ or $a_i^{\leftarrow}$, and every Hamiltonian $s$-$t$-path uses $a_i^{\rightarrow}$ if it traverses the chain from left to right and $a_i^{\leftarrow}$ if it traverses it from right to left.
Hence, with $f(\ell_i) = a_i^{\rightarrow}$ and $f(\overline\ell_i) = a_i^{\leftarrow}$, a literal belongs to a satisfying assignment if and only if its image belongs to the corresponding Hamiltonian path.

This specific correspondence motivates a stricter goal for reductions between problems in $\NPS$:
mapping the underlying solution structure of the first problem directly into the other.
Suppose that $\Pi = (\I,\U,\sol)$ and $\Pi' = (\I',\U',\sol')$ are problems in $\NPS$, and that $g$ is a reduction from $\Pi$ to $\Pi'$.
Let $I \in \I$ be an instance of $\Pi$ and $f: \U(I) \rightarrow \U'(g(I))$ be a function.
We define the \emph{embedded universe} by $\Uemb'(I) := f(\U(I))$ and may write $\Uemb'$ if $I$ is clear from the context.
We say that $f$ is \emph{solution-embedding} if it satisfies the following three properties:
\label{def:se-function}

\begin{enumerate}
    \item The function $f$ is injective,
    \item $\forall S \in \sol(I) \ 
            \exists S' \in \sol'(g(I)): f(S) = S' \cap \U'_{emb}$,
    \item   $\forall S' \in \sol'(g(I)) :  
             f^{-1}(S'\cap \U'_{emb}) \in \sol(I)$.
\end{enumerate}

The following lemma gives an equivalent formulation, which is often more convenient.

\begin{lemma}
\label{lem:se-set-equality}
Let $f\colon \U(I) \to \U'(g(I))$ be injective, and let $\U'_{emb} = f(\U(I))$.
Then $f$ is solution-embedding if and only if
$$
    \{f(S) : S \in \sol(I)\}  =\{S' \cap \U'_{emb} : S' \in \sol'(g(I))\}.
$$
\end{lemma}
\begin{proof}
    Suppose that $f$ is solution-embedding.
    Property~2 gives the inclusion ``$\subseteq$''.
    For ``$\supseteq$'', let $S' \in \sol'(g(I))$.
    Since $S' \cap \U'_{emb} \subseteq f(\U(I))$, we have $S' \cap \U'_{emb} = f(f^{-1}(S' \cap \U'_{emb}))$, and $f^{-1}(S' \cap \U'_{emb}) \in \sol(I)$ by property~3.

    Conversely, suppose that the equality holds.
    Property~1 holds by assumption, and property~2 is immediate.
    For property~3, let $S' \in \sol'(g(I))$.
    Then $S' \cap \U'_{emb} = f(S)$ for some $S \in \sol(I)$, and hence $f^{-1}(S' \cap \U'_{emb}) = f^{-1}(f(S)) = S \in \sol(I)$ by injectivity.
\end{proof}

\begin{definition}[Solution-embedding Reduction, from \cite{DBLP:journals/corr/abs-2602-12350}]
\label{def:se-reduction}
    Let $\Pi = (\I,\U,\sol)$ and $\Pi' = (\I',\U',\sol')$ be two problems in $\NPS$.
    We say that there is a solution-embedding reduction from $\Pi$ to $\Pi'$, and write $\Pi \leqSE \Pi'$, if
    \begin{itemize}
        \item There exists a function $g : \{0,1\}^* \to \{0,1\}^*$ computable in polynomial time in the input size $|I|$, such that $\sol(I) \neq \emptyset$ iff $\sol'(g(I)) \neq \emptyset$.
        \item For all $I \in \I$, there exists a function $f_I$ uniformly computable in polynomial time in $|I|$ such that $f_I$ is a solution-embedding mapping from $\U(I)$ to $\U'(g(I))$.
    \end{itemize}
\end{definition}

In this sense, an SE reduction preserves the solution structure in both directions.
Every source solution is represented by the embedded part of some target solution, and every target solution induces a source solution.
The mapping $f_I$ therefore preserves the correspondence between individual solution elements rather than merely preserving the existence of a solution.
We remark that the first condition of \Cref{def:se-reduction} is implied by the second one:
by property~2, $\sol(I) \neq \emptyset$ implies $\sol'(g(I)) \neq \emptyset$, and by property~3, the converse holds as well.

\subparagraph*{$\NPS$-Completeness.}

This stronger notion of reduction is used as the basis for the definition of  $\NPS$-completeness. 
A problem is \emph{$\NPS$-complete} if it belongs to $\NPS$ and every problem in $\NPS$ can be reduced to it by a solution-embedding reduction.
The authors of \cite{DBLP:journals/corr/abs-2602-12350} show that Cook's proof that \textsc{Satisfiability} is $\NP$-complete can be refined to show that \textsc{Satisfiability}, lifted to a problem in $\NPS$, is $\NPS$-complete.
They also show that many classical $\NP$-complete problems, including \textsc{Clique}, \textsc{Vertex Cover}, \textsc{Knapsack}, and \textsc{Traveling Salesman}, are $\NPS$-complete.
Subsequently, the authors of \cite{DBLP:journals/corr/abs-2602-12350} describe several natural multilevel extensions of problems in $\NPS$ to problems that are at some level (e.g., $\Sigma_2^p$) of the polynomial hierarchy.
Then, they prove that if the original problem is $\NPS$-complete, then the corresponding multilevel problem is complete for its level in the polynomial hierarchy.

\subsection{Linear Objective Problems}
\label{sec:lop}

We next consider optimization problems in which the goal is to select a feasible solution that minimizes (or maximizes) some linear objective function.
If phrased as a decision problem, we typically want to determine whether there is a feasible solution that achieves some threshold value.

More precisely, let $\Pi=(\I,\U,\sol)$ be an $\NPS$ problem.
For an instance $I \in \I$, we equip it with a weight function $w:\mathcal U(I)\rightarrow\mathbb Z_{\geq 0}$.
The weight of a solution $S\subseteq\mathcal U(I)$ is defined by $w(S) = \sum_{u\in S} w(u)$.
A linear objective problem thus consists of all instances of $\Pi$ paired with a weight function.
We consider both minimization and maximization problems.

\subparagraph*{Non-negative weights.}
Throughout, we restrict weights to be non-negative integers.
This is the natural setting for the classical problems we consider: for instance, a vertex of negative weight would always be included in a minimum weight vertex cover, and \textsc{Knapsack} with negative item values is not meaningful.
Moreover, the restriction only strengthens our hardness results, since hardness for non-negative weights implies hardness for arbitrary integral weights.
The restriction does mean, however, that a maximization problem cannot simply be turned into a minimization problem by negating its objective.
In \Cref{sec:regret-max}, we replace negation by a notion of duality based on complementation, which preserves non-negativity.

This leads to the following notion.

\begin{definition}[Linear Objective Problem]
\label{def:lop-opt}
    A \emph{linear objective problem} $\Lambda_{\textsc{opt}} = (\I, \U, \sol, w, \sigma)$ based on an $\NPS$ problem $\Pi = (\I_0, \U, \sol)$ consists of
    \begin{itemize}
        \item $\I \subseteq \{0,1\}^*$ is a language. We call $\I$ the set of instances of $\Lambda_{\textsc{opt}}$.
        Every instance $I \in \I$ is a tuple $I = (I_0, w)$ of an underlying instance $I_0 \in \I_0$ and a weight function $w^{(I)}: \U(I) \to \Z_{\geq0}$.
        \item To each instance $I \in \I$, we associate the universe $\U(I) = \U(I_0)$.
        \item To each instance $I \in \I$, we associate the solution set $\sol(I) = \sol(I_0)$.
        \item $\sigma \in \{\min, \max\}$ is the optimization sense.
    \end{itemize}

    The set $\sol$ refers to all feasible solutions of the $\LOP$ problem.
    We let $\sol_{opt}$ denote the set of optimal solutions, i.e. $\sol_{opt}(I) = \arg\sigma_{S \in \sol(I)} w^{(I)}(S)$.
    If $\sol(I)=\emptyset$, then $\sol_{opt}(I) = \emptyset$.
\end{definition}

As an example, let us consider vertex cover problems.
We start with the decision problem \textsc{Feasible Vertex Cover}, which can
be expressed as an $\NPS$ problem $(\I,\U,\sol)$ as follows.
The instance set $\I$ consists of all graphs.
For each instance $I \in \I$, the universe $\U(I)$ is the set of vertices, and the solution set $\sol(I)$ is the set of all vertex covers of the graph in $I$.
This vertex cover variant is easy to solve since the set of all vertices is always a vertex cover.
Based on this $\NPS$ problem, we now define the $\LOP$ problem \textsc{Minimum Weight Vertex Cover} $ = (\I,\U,\sol,w,\min)$.
That is, we add the optimization sense $\min$ and associate a weight function $w^{(I)}$ to each instance $I \in \I$.
In the case that $w^{(I)}$ is always the vector of 1's, the problem is \textsc{Minimum Cardinality Vertex Cover}.
In comparison to the $\NPS$ version, the $\LOP$ version \textsc{Minimum Weight Vertex Cover} is $\NP$-hard to solve.

In \Cref{sec:LOPcompleteness}, we will define $\LOP$-completeness in a way that parallels the definition of $\NPS$-completeness.
Recall that a problem $\Pi \in \NPS$ is $\NPS$-complete if for every other problem $\Pi' \in \NPS$, there is an SE reduction from $\Pi'$ to $\Pi$.
Moreover, the SE reduction establishes a mapping between each set of feasible solutions $\sol'$ of $\Pi'$ and the corresponding set of feasible solutions $\sol$ of $\Pi$.
For $\LOP$-completeness, we will say that a problem $\Lambda \in \LOP$ is $\LOP$-complete if for every other problem $\Lambda' \in \LOP$, there is a special type of SE reduction from $\Lambda'$ to $\Lambda$.
This special type of SE reduction establishes a mapping between each set of optimal solutions $\sol'_{opt}$ of $\Lambda'$ and the corresponding set of optimal solutions $\sol_{opt}$ of $\Lambda$.

These reductions will be computable in polynomial time, even though deciding whether a given feasible solution is optimal is a $\coNP$ problem and, for many problems, $\coNP$-complete.
That is, we will construct polynomial-time reductions that map optimal solutions to optimal solutions without ever having to recognize an optimal solution.

\subsubsection{Decision Version of LOP}
The complexity class $\LOP$ consists of the decision problems obtained from linear objective problems by adding a threshold.
When we refer to the problem of \Cref{def:lop-opt} itself, we use the term ``$\LOP$ optimization problem'' or ``optimization version''.
When discussing reductions and completeness, we work with the decision version, which we define as follows.

\begin{definition}[Linear Objective Problem (Decision Version)]
\label{def:LOP}
    A problem $\Lambda_{\textsc{dec}} = (\I, \U, \sol, w, t, \sigma)$ is in $\LOP$ if there is an $\LOP$ optimization problem $\Lambda_{\textsc{opt}} = (\I_{\textsc{opt}}, \U, \sol, w, \sigma)$ based on some $\Pi = (\I_0, \U, \sol) \in \NPS$ such that
    \begin{itemize}
        \item every instance $I \in \I \subseteq \set{0,1}^*$ is a tuple $I = (I_0,w,t)$ of an instance $(I_0, w) \in \I_{\textsc{opt}}$ and a threshold $t^{(I)} \in \mathbb Z$, and
        \item the universe, the solution set, the weight function, and the optimal solutions of $I$ are those of the instance $(I_0, w)$ of $\Lambda_{\textsc{opt}}$.
    \end{itemize}
    The set $\sol$ refers to all feasible solutions of the linear objective problem.
    For a \emph{minimization $\LOP$} problem ($\sigma = \min$), we define the set of \emph{threshold-feasible} solutions by $\sol_t(I) = \{S \in \sol(I) : w^{(I)}(S) \leq t^{(I)}\}$.
    For a \emph{maximization $\LOP$} problem ($\sigma = \max$), $\sol_t(I) = \{S \in \sol(I) : w^{(I)}(S) \geq t^{(I)}\}$.
    The question associated with an instance $I$ is whether $\sol_t(I) \neq \emptyset$.
\end{definition}

In the following, we write $\sol_{z}(I)$, $z \in \Z$, for the threshold-feasible solutions of $I$ when its threshold $t^{(I)}$ is replaced by $z$, i.e. $\sol_z(I) = \{S \in \sol(I) : w^{(I)}(S) \leq z\}$ for $\sigma = \min$ and $\sol_z(I) = \{S \in \sol(I) : w^{(I)}(S) \geq z\}$ for $\sigma = \max$.

We allow negative thresholds.
This is harmless:
a minimization instance with a negative threshold is a trivial no-instance, and a maximization instance with a negative threshold is equivalent to the same instance with threshold $0$.
Negative thresholds may arise when taking duals in \Cref{sec:regret-max}.

\subparagraph*{Notation for instances.}
We write $I$ for an instance of the $\LOP$ decision problem, and $w^{(I)}$ and $t^{(I)}$ for the weight function and threshold encoded in it.
Since the universe and the solution set are those of the corresponding $\NPS$ problem $\Pi$, the size of the universe is polynomial in the input, and the verifier of $\Pi$ decides membership in $\sol(I)$ in polynomial time.
When we want to vary the threshold of a fixed instance of the $\LOP$ optimization problem $\Lambda_{opt}$, we write $(I, t)$ for the decision instance obtained from an instance $I$ of $\Lambda_{opt}$ by adding the threshold $t$.
Given a problem $\Pi \in \NPS$, we also refer to the corresponding linear objective optimization and decision problems as $\Lambda_{opt} = (\Pi, w, \sigma)$ and $\Lambda_{dec} = (\Pi, w, t, \sigma)$, respectively.

Let us consider once more the problem \textsc{Vertex Cover} as an example.
Again, let $(\I,\U,\sol)$ be the $\NPS$ problem, where for each instance $I \in \I$, the universe $\U(I)$ is the set of vertices and the solution set $\sol(I)$ is the set of all vertex covers.
The corresponding decision version of \textsc{Minimum Weight Vertex Cover} is the $\LOP$ decision problem $(\I, \U, \sol, w, t, \min)$ that includes an additional threshold $t$ for every instance $I \in \I$.
We, therefore, ask the question: Does the minimum weight vertex cover in instance $I$ have weight at most $t^{(I)}$?

\subparagraph*{The relation of $\LOP$ optimization and decision problems.}
The decision problems in $\LOP$ retain the optimal solutions of the corresponding optimization problems.
Let $I$ be an instance of $\Lambda_{opt}$ with $\sol(I) \neq \emptyset$ and let $z^*$ be its optimal objective value.
We define $\sol_{opt}(I,t) = \sol_{opt}(I)$ for every threshold $t$.
If $t = z^*$, then the threshold-feasible solutions are precisely the optimal ones, i.e. $\sol_{t}(I,t) = \sol_{opt}(I,t)$.
Consequently, if a solution-embedding reduction between two $\LOP$ decision problems can guarantee that the optimal value is used as threshold, then optimal solutions of the first problem are mapped to optimal solutions of the second problem.
We formalize this in \Cref{sec:LOPcompleteness}.

\subsubsection{The relation between $\LOP$, $\NPS$, and $\NP$}
We next discuss the relation between the complexity classes $\LOP$, $\NPS$, and $\NP$.
The purpose of this subsection is to make explicit what information is retained when an $\LOP$ problem is viewed as an $\NPS$ problem or as an ordinary $\NP$ language.

An $\LOP$ problem contains two kinds of information.
First, it specifies a set of feasible solutions through its underlying $\NPS$ problem.
Second, it specifies an objective function and a threshold, which select a subset of these feasible solutions.
When an $\LOP$ decision problem is viewed as an $\NPS$ problem, the objective and threshold can be incorporated into the solution set.
This transformation preserves the threshold-feasible solutions, but it no longer distinguishes the objective and threshold from the underlying feasibility constraints.

The following lemma formalizes this transformation.

\begin{lemma}
\label{lem:lop-to-nps}
     Each problem $\Lambda = (\I, \U, \sol, w, t,\sigma)$ in $\LOP$ can be transformed into a problem  $\Pi = (\I_{\Pi}, \U_{\Pi}, \sol_\Pi)$ in $\NPS$ with the same instances and universes and with solution sets $\sol_\Pi(I) = \sol_t(I)$.
\end{lemma}
\begin{proof}
    Let $\Lambda = (\I, \U, \sol, w, t,\sigma)$ be in $\LOP$.
    We construct the problem $\Pi = (\I_{\Pi}, \U_{\Pi}, \sol_\Pi)$ such that $\Pi$ is the $\NPS$ problem derived from the $\LOP$ problem $\Lambda$.

    Since the weight function and threshold are part of the input word of an $\LOP$ instance, we use the same input words for $\Pi$ and define $\I_\Pi=\I$.
    Moreover, for every $I \in \I$, we define $\U_\Pi(I)=\U(I)$.
    Furthermore, let $\sol_\Pi(I) = \sol_{t}(I).$
    The solution set can be verified in polynomial time:
    the verifier first checks whether $S \in \sol(I)$ and then evaluates the linear objective $w^{(I)}(S)$ and checks the corresponding threshold condition: $w^{(I)}(S) \leq t^{(I)}$ for minimization or $w^{(I)}(S) \geq t^{(I)}$ for maximization.
    Hence $\Pi \in \NPS$.
\end{proof}

The transformation has an important consequence.
The information contained in $w^{(I)}$ and $t^{(I)}$ is absorbed into the new solution set $\sol_\Pi(I) = \sol_{t}(I)$.
The resulting $\NPS$ problem retains exactly the threshold-feasible solutions, but no longer distinguishes whether these solutions arose from the underlying feasibility constraints or from imposing the particular objective and threshold.
If we subsequently equip the derived $\NPS$ problem with a new objective function and threshold, these new parameters are independent of the objective and threshold of the original $\LOP$ problem.

For example, the $\NPS$ problem derived from the $\LOP$ problem \textsc{Minimum Weight Vertex Cover} has, for an instance $(G, w, t)$, the solution set
$$
    \sol_t(I) = \{S \subseteq V : S \text{ is a vertex cover and } w(S) \leq t\}.
$$
In this derived problem, the distinction between the vertex-cover constraint and the weight constraint has disappeared:
both are simply conditions checked by the verifier.
If we now equip the derived problem with a second weight function $w'$ and threshold $t'$, the resulting $\LOP$ problem asks whether there is a set $S$ satisfying
$$
    S \text{ is a vertex cover}, \quad w(S) \leq t, \quad w'(S) \leq t'.
$$
The original weight function $w$ has therefore changed its semantic role:
it is no longer an objective function of the resulting $\LOP$ problem, but part of the feasibility predicate inherited from the derived $\NPS$ problem.
This illustrates why the passage from $\LOP$ to $\NPS$ preserves the relevant solution set but does not preserve the decomposition of the problem into feasibility and optimization.

\subparagraph{The relation to $\NP$.}
Finally, we show that the problems in $\LOP$ can be transformed into a language in $\NP$ by encoding the instance correspondingly and using the verifier to implicitly define the universe and solution set.

\begin{lemma}
\label{lem:lop-to-np}
    Each problem $\Lambda$ in $\LOP$ can be transformed into a language $L_\Lambda$ in $\NP$.
\end{lemma}
\begin{proof}
    By \Cref{lem:lop-to-nps}, $\Lambda$ can be transformed into a problem $\Pi \in \NPS$ with $\sol_\Pi(I) = \sol_t(I)$, and by \Cref{lem:subsetNP}, $\Pi$ can be transformed into a language in $\NP$.
\end{proof}

\subsubsection{$\LOP$-Completeness}
\label{sec:LOPcompleteness}
We now introduce a completeness notion for $\LOP$ problems.
Roughly speaking, we will say that a problem $\Lambda \in \LOP$ is $\LOP$-complete if for every other problem $\Lambda' \in \LOP$ there is a reduction from $\Lambda'$ to $\Lambda$ such that optimal solutions in $\Lambda'$ correspond to optimal solutions in $\Lambda$.
Thus, in an important technical sense, the $\LOP$-complete problem $\Lambda$ can represent the optimal solution structure of every optimization problem in $\LOP$.

In \Cref{sec:regret-min}, we will apply this concept to bilevel optimization problems.
A follower's problem in bilevel optimization can usually be represented as a problem in $\LOP$.
In \Cref{sec:regret-min,sec:regret-max}, we obtain results showing that a certain subarea of bilevel optimization, namely min-max regret optimization, is $\Sigma_2^p$-complete whenever the follower's problem is $\LOP$-complete.

\subparagraph*{An example.}
We illustrate the idea with a folklore reduction from \textsc{Minimum Cardinality Vertex Cover} to \textsc{Minimum Cardinality Dominating Set}.
In both problems, an instance is a graph $G=(V,E)$ together with unit weights $w \equiv 1$ and a threshold $t$, and the universe is the vertex set $V$.
For \textsc{Minimum Cardinality Vertex Cover}, the solution set consists of all vertex covers of $G$, i.e. all sets $S$ such that at least one endpoint of each edge $e \in E$ is in $S$.
For \textsc{Minimum Cardinality Dominating Set}, the solution set consists of all dominating sets of $G$, i.e. all sets $S$ such that each vertex is in $S$ or has a neighbor in $S$.
For simplicity, we restrict the vertex cover problem to graphs without isolated vertices; an isolated vertex never belongs to a minimum vertex cover, but it would have to be dominated.

Let $(G, w, t)$ be an instance of \textsc{Minimum Cardinality Vertex Cover}.
We construct the graph $G' = (V', E')$ from $G$ by adding, for each edge $\{u,v\} \in E$, two new vertices $uv_1$ and $uv_2$ together with the edges $\{u,uv_i\}$ and $\{v,uv_i\}$ for $i \in \{1,2\}$.
Thus $V' = V \cup \{uv_1, uv_2 : \{u,v\} \in E\}$ and $E \subseteq E'$.
All vertices of $G'$ receive weight $1$, the threshold is $t' = t$, and the embedding $f$ is the identity on $V$.
We first relate the two solution sets.

\begin{itemize}
    \item \emph{Every vertex cover $S$ of $G$ is a dominating set of $G'$.}
    A gadget vertex $uv_i$ is dominated since $u \in S$ or $v \in S$. A vertex $x \in V \setminus S$ has an incident edge $\{x,y\} \in E$, since $G$ has no isolated vertices, and $y \in S$ because $S$ is a vertex cover.

    \item \emph{For every dominating set $D$ of $G'$, there is a vertex cover $S \subseteq V$ of $G$ with $|S| \leq |D|$, and $|S| < |D|$ if $D \not\subseteq V$.}
    Let $S$ consist of $D \cap V$ together with one arbitrary endpoint of each edge $\{u,v\} \in E$ with $D \cap \{u,v\} = \emptyset$.
    Clearly, $S$ is a vertex cover.
    If $D \cap \{u,v\} = \emptyset$, then both $uv_1$ and $uv_2$ belong to $D$, since the closed neighborhood of $uv_i$ is $\{uv_i, u, v\}$.
    As the gadget vertices of different edges are distinct, at most $|D \setminus V|/2$ vertices are added, so $|S| \leq |D \cap V| + |D \setminus V|/2$, which is less than $|D|$ if $D \setminus V \neq \emptyset$.
\end{itemize}

Consequently, $G$ has a vertex cover of size at most $t$ if and only if $G'$ has a dominating set of size at most $t'$, so the construction is a many-one reduction.
Moreover, the minimum vertex cover size of $G$ equals the minimum dominating set size of $G'$; call it $k^*$.
Every minimum dominating set $D$ of $G'$ is contained in $V$, as otherwise the second statement yields a vertex cover of size less than $k^*$.
For such $D$, no vertex is added in the construction of $S$, because an edge with $D \cap \{u,v\} = \emptyset$ would force gadget vertices into $D$; hence $D$ itself is a minimum vertex cover of $G$.
Conversely, by the first statement, every minimum vertex cover of $G$ is a dominating set of $G'$ of size $k^*$.
Therefore,
$$
    \{S : S \text{ is a min. card. vertex cover in } G\} = \{D \cap V : D \text{ is a min. card. dom. set in } G'\}.
$$
In particular, the threshold $t'$ is tight for $G'$ (it equals $k^*$) if $t$ is tight for $G$, and in this case the identity is solution-embedding from the threshold-feasible solutions of $G$ to those of $G'$.

The reduction does \emph{not}, however, preserve the solution structure for non-tight thresholds.
If $G$ consists of a single edge $\{u,v\}$ and $t = 2$, then $D = \{uv_1, uv_2\}$ is a dominating set of $G'$ of size $2 \leq t'$, but $D \cap V = \emptyset$ is not a vertex cover.
This is typical: many natural reductions preserve the structure of optimal solutions but not that of all threshold-feasible solutions.
The following definitions are designed to capture exactly this behavior.

\subparagraph*{Tight thresholds and tight reductions.}
The definition of $\LOP$-completeness relies on the concept of a ``tight threshold'' for an $\LOP$ decision problem. 

\begin{definition}[Tight Threshold]
\label{def:tight}
    Let $\Lambda = (\I,\U,\sol,w,t,\sigma)$ be in $\LOP$.
    When $\sigma = \min$, the threshold $t^{(I)} =: t$ is \emph{tight} if $\sol_t(I) \neq \emptyset$ and $\sol_{t-1}(I) = \emptyset$.
    When $\sigma = \max$, it is tight if $\sol_t(I) \neq \emptyset$ and $\sol_{t+1}(I) = \emptyset$.
\end{definition}

We call an instance $I \in \I$ tight if its threshold $t^{(I)}$ is tight.
A no-instance, i.e. an instance with $\sol_t(I) = \emptyset$, in particular one with $\sol(I) = \emptyset$, is never tight.
For a yes-instance, the optimal value $z^*$ is the unique tight threshold, and on a tight instance the threshold-feasible solutions are exactly the optimal ones, which we prove in the following lemma.
For example, consider \textsc{Satisfiability (Literal Universe)} with unit weights:
every feasible solution has exactly $|\U(I)|/2$ literals, so for a satisfiable instance the unique tight threshold is $t = |\U(I)|/2$, and then $\sol_t(I) = \sol(I) = \sol_{opt}(I)$.

\begin{lemma}
\label{lem:tight-characterization}
    Let $\Lambda = (\I,\U,\sol,w,t,\sigma)$ be in $\LOP$, let $I \in \I$ with $\sol(I) \neq \emptyset$.
    \begin{enumerate}
        \item When $\sigma = \min$, $t^{(I)}$ is tight if and only if $t^{(I)} = \min_{S \in \sol(I)} w^{(I)}(S)$
        \item When $\sigma = \max$, $t^{(I)}$ is tight if and only if $t^{(I)} = \max_{S \in \sol(I)} w^{(I)}(S)$
        \item If $t^{(I)}$ is tight, then $\sol_t(I) = \sol_{opt}(I)$
    \end{enumerate}
\end{lemma}
\begin{proof}
    Let $\sigma = \min$ and $z^* = \min_{S \in \sol(I)} w^{(I)}(S)$, which exists since $\sol(I) \neq \emptyset$; the case $\sigma = \max$ is symmetric.
    Since all weights are integers, $\sol_t(I) \neq \emptyset$ holds if and only if $z^* \leq t^{(I)}$, and $\sol_{t-1}(I) = \emptyset$ holds if and only if $z^* > t^{(I)}-1$, i.e. $z^* \geq t^{(I)}$.
    Hence $t^{(I)}$ is tight if and only if $t^{(I)} = z^*$, which proves Statements 1 and 2.
    In this case, $\sol_t(I) = \{S \in \sol(I) : w^{(I)}(S) \leq z^*\} = \sol_{opt}(I)$, which proves Statement 3.
\end{proof}

We are now ready to define a reduction that preserves both solution structure and tightness.
The solution-embedding property is required only for tight instances; on such instances, threshold-feasible and optimal solutions coincide.

\begin{definition}[Tight Solution-embedding Reduction]
\label{def:tight-se-reduction}
    Let $\Lambda = (\I,\U,\sol,w,t,\sigma)$ and $\Lambda' = (\I',\U',\sol',w',t',\sigma')$ be two problems in $\LOP$.
    We say that there is a tight solution-embedding reduction from $\Lambda$ to $\Lambda'$, and write $\Lambda \leqTSE \Lambda'$, if
    \begin{description}
        \item[(1)] there exists a function $g : \{0,1\}^* \to \{0,1\}^*$ computable in polynomial time in the input size $|I|$, such that $\sol_t(I) \neq \emptyset$ iff $\sol'_{t'}(g(I)) \neq \emptyset$ for every $I \in \I$;
        \begin{description}
            \item[(a)] $g(I)$ is tight for every tight instance $I \in \I$;
        \end{description}
        \item[(2)] there exists a family of functions $f_I: \U(I) \to \U'(g(I))$, uniformly computable in time polynomial in $|I|$, such that
        \begin{description}
            \item[(a)] $f_I$ is injective for every $I \in \I$;
            \item[(b)] $f_I$ is solution-embedding for every $I \in \I$ that is tight.
        \end{description}
    \end{description}
\end{definition}

A tight solution-embedding reduction is a many-one reduction on all instances by property $(1)$ and behaves like a solution-embedding reduction on all no-instances and tight yes-instances.
If $I$ is a no-instance, then $\sol_t(I) = \emptyset$ and $I$ is not tight.
By property (1) $\sol'_{t'}(g(I)) = \emptyset$, so property (2) only asks that $f_I$ be injective.
This injectivity of $f_I$ is usually satisfied for reductions.
For a correctly encoded instance $I \in \I$, the universe $\U(g(I))$ of the target instance $g(I)$ needs to include $f(\U(I))$.
A typical reduction does exactly that.
For instance, consider the reduction from \textsc{Vertex Cover} to \textsc{Dominating Set} above.
In this reduction, we introduce a vertex in the \textsc{Dominating Set} for every vertex in the \textsc{Vertex Cover} instance.
This is the usual strategy to produce instances in such a way that in the correctness proof, one is able to prove a one-to-one correspondence between the elements in a solution.
(For incorrectly encoded words, we might assume that the universe is empty. Note that any function with an empty domain is injective.)
Since for a no-instance, we have $\sol_t(I) = \emptyset$, properties 2 and 3 of a solution-embedding function are trivially fulfilled for the empty set.
Thus, if $f_I$ is injective and $\sol_t(I) = \emptyset$, then $f_I$ is solution-embedding.

Now, consider a yes-instance $I$.
If $I$ is a tight instance, then the solution-embedding properties need to be fulfilled.
Furthermore, the target instance $g(I)$ needs to be tight as well.
We, therefore, obtain a mapping $f_I$ that characterizes the solution-embedding between optimal solutions $\sol_{opt}(I)$ in $I$ and optimal solutions $\sol'_{opt}(g(I))$ in $g(I)$.
If $I$ is not tight but $\sol_t(I) \neq \emptyset$, only injectivity is required of $f_I$.
This is the freedom used in the reduction from \textsc{Vertex Cover} to \textsc{Dominating Set} above.
All reductions in this paper satisfy these properties.

In contrast to \Cref{def:se-reduction}, a tight solution-embedding reduction does not need to be solution-embedding for all thresholds but only for tight ones.
Hence the solution-embedding property of a TSE reduction is precisely a correspondence between optimal solutions.
The example from above from \textsc{Vertex Cover} to \textsc{Dominating Set} is a tight solution-embedding reduction, even though it is not solution-embedding for non-tight thresholds.

Applying the solution-embedding properties to these two sets yields the following lemma.

\begin{lemma}
\label{lem:tse-optimal}
    Let $\Lambda = (\I,\U,\sol,w,t,\sigma)$ and $\Lambda' = (\I',\U',\sol',w',t',\sigma')$ be two problems in $\LOP$.
    If there is a tight solution-embedding reduction $(g,f_I)$ from $\Lambda$ to $\Lambda'$,
    then for every tight instance $I \in \I$, with $\U'_{emb} = f_I(\U(I))$,
    $$
        \{f_I(S): S \in \sol_{opt}(I)\} = \{S' \cap \U'_{emb} : S' \in \sol'_{opt}(g(I))\}
    $$
    and
    $$
        \sol_{opt}(I) = \{f_I^{-1}(S' \cap \U'_{emb}): S' \in \sol'_{opt}(g(I))\}.
    $$
\end{lemma}
\begin{proof}
    Since $I$ is tight, $\sol_t(I) = \sol_{opt}(I)$ by \Cref{lem:tight-characterization}.
    By \Cref{def:tight-se-reduction}, $g(I)$ is tight, and hence $\sol'_{t'}(g(I)) = \sol'_{opt}(g(I))$ by \Cref{lem:tight-characterization}.
    The first equality now follows from \Cref{lem:se-set-equality}, since $f_I$ is solution-embedding from $\sol_t(I)$ to $\sol'_{t'}(g(I))$.
    The second equality follows from the first one and the injectivity of $f_I$.
\end{proof}

Thus, tight solution-embedding reductions preserve optimal solution structure whenever the source threshold is tight.
This is the central property of tight solution-embedding reductions that we will use below.

\begin{lemma}
\label{lem:TSE-transitive}
    Tight solution-embedding reductions are transitive.
\end{lemma}
\begin{proof}
    Let $(g_1, f_1)$ be a tight solution-embedding reduction from $\Lambda_1$ to $\Lambda_2$, and let $(g_2, f_2)$ be one from $\Lambda_2$ to $\Lambda_3$. For $i \in \{1,2,3\}$, we write $\U_i$ and $\sol_{t,i}$ for the universe and the set of threshold-feasible solutions of an instance of $\Lambda_i$. As in \Cref{def:tight-se-reduction}, solution-embedding refers to the sets of threshold-feasible solutions.
    \begin{description}
        \item[The composed reduction.]
        Define $g := g_2 \circ g_1$.
        For an instance $I$ of $\Lambda_1$, let $I_2 := g_1(I)$ and $I_3 := g_2(I_2) = g(I)$, and define
        $$
            f_I := (f_2)_{I_2} \circ (f_1)_I \colon \U_1(I) \to \U_3(I_3).
        $$
        Since $g_1$ runs in polynomial time, $|I_2|$ is polynomial in $|I|$.
        Hence $g$ is computable in time polynomial in $|I|$.
        Likewise, $f_I$ is computable uniformly in time polynomial in $|I|$, because $(f_1)_I$ is computable in time polynomial in $|I|$ and $(f_2)_{I_2}$ in time polynomial in $|I_2|$.
        Moreover, $f_I$ is injective for every $I$, as a composition of injective maps.
        
        \item[Many-one property.]
        For every instance $I$ of $\Lambda_1$, the first condition of \Cref{def:tight-se-reduction} for $(g_1,f_1)$ and for $(g_2,f_2)$ gives
        $$
            \sol_{t,1}(I) \neq \emptyset
            \iff \sol_{g_1(t),2}(I_2) \neq \emptyset
            \iff \sol_{g_2 \circ g_1 (t),3}(I_3) \neq \emptyset.
        $$
        
        \item[Tightness.]
        Let $I$ be a tight instance of $\Lambda_1$.
        Since $(g_1,f_1)$ is a tight solution-embedding reduction, $I_2$ is tight.
        Since $(g_2,f_2)$ is a tight solution-embedding reduction and $I_2$ is tight, $I_3 = g(I)$ is tight.
    
        \item[Solution embedding.]
        Let $I$ be tight.
        The function $f_I$ is injective and maps to $\U_3(I_3)$ and is polynomial-time computable.
        It remains to show that $f$ has the desired solution-embedding property matching \Cref{lem:se-set-equality}.
        In order to reduce the notation, we omit the subscript in $f$ (i.e. $f = f_2 \circ f_1$).
        We also write $\U_1, \U_2, \U_3$ instead of $\U_1(I_1), \U_2(I_2), \U_3(I_3)$ and $\sol_{t_1,1},\sol_{t_2,2}, \sol_{t_3,3}$ instead of  $\sol_{t,1}(I_1),\sol_{g_1(t),2}(I_2), \sol_{g_2\circ g_1(t),3}(I_3)$.
        Since $\Lambda_1 \leqTSE \Lambda_2$ and $\Lambda_2 \leqTSE \Lambda_3$ and since for injective functions it holds that $f(A \cap B) = f(A) \cap f(B)$, we have
        \begin{align*}
            \set{f(S_1) : S_1 \in \sol_{t_1,1}} = &\ \set{f_2(f_1(S_1)) : S_1 \in \sol_{t_1,1}}\\
            = &\ \set{f_2(Y) : Y \in \set{f_1(S_1) : S_1 \in \sol_{t_1,1}}}\\
            = &\ \set{f_2(Y) : Y \in \set{S_2 \cap f_1(\U_1) : S_2 \in \sol_{t_2,2}}}\\
            = &\ \set{f_2(S_2) \cap f_2(f_1(\U_1)) : S_2 \in \sol_{t_2,2}}\\
            = &\ \set{S_3 \cap f_2(\U_2) \cap f_2(f_1(\U_1)) : S_3 \in \sol_{t_3,3}}\\
            = &\ \set{S_3 \cap f(\U_1) : S_3 \in \sol_{t_3,3}}.
        \end{align*}
    \end{description}
\end{proof}

\subparagraph*{$\LOP$-completeness.}

We say that a problem $\Lambda \in \LOP$ is \emph{$\LOP$-complete} if $\Lambda' \leqTSE \Lambda$ for every $\Lambda' \in \LOP$.

We can additionally relate $\NPS$-completeness and $\LOP$-completeness.
For an NP-S problem $\Pi$, the \emph{zero-weight LOP problem derived from $\Pi$} equips every instance $I_0$ of $\Pi$ with the weight function $w \equiv 0$ and the threshold $t = 0$.
Every solution then has weight $0$, so $\sol_t(I) = \sol(I) = \sol_{opt}(I)$, and every yes-instance is tight, since $\sol_{-1}(I) = \emptyset$ for $\sigma = \min$ and $\sol_{1}(I) = \emptyset$ for $\sigma = \max$.
In particular, the optimization sense of a zero-weight problem is irrelevant:
the minimization and the maximization version have the same instances, the same threshold-feasible solutions, and the same tight instances, so the identity is a tight solution-embedding reduction between them in both directions.
We therefore speak of \emph{the} zero-weight LOP problem derived from $\Pi$, without specifying its sense.

\begin{lemma}\label{lem:NP-SandLOP-completeness}
    Let $\Pi \in \NPS$, and let $\Lambda$ be the zero-weight $\LOP$ problem derived from $\Pi$.
    Then $\Pi$ is $\NPS$-complete if and only if $\Lambda$ is $\LOP$-complete.
\end{lemma}
\begin{proof}
    For the forward direction, let $\Pi = (\I,\U,\sol)$ be $\NPS$-complete, and let $\Lambda_0 \in \LOP$ be arbitrary.
    Let $\Pi_0$ be the $\NPS$ problem derived from $\Lambda_0$ by \Cref{lem:lop-to-nps}, whose solution sets are $\sol_{t_0}(I)$.
    Since $\Pi$ is $\NPS$-complete, there is an SE reduction $(g_0, f)$ from $\Pi_0$ to $\Pi$.
    Let $I_0$ be an instance of $\Pi_0$ and $g(I_0) := (g_0(I_0), 0, 0)$ be the corresponding instance of $\Lambda$.
    Since $\sol_t(g(I_0)) = \sol(g_0(I_0))$, we have $\sol_{t,0}(I_0) \neq \emptyset$ iff $\sol_t(g(I_0)) \neq \emptyset$, and $f_{I_0}$ is solution-embedding for every instance $I_0$, in particular for every tight one.
    Furthermore, if $I_0$ is tight, then $\sol_{t,0}(I_0) \neq \emptyset$, so $g(I_0)$ is a yes-instance.
    Finally, every yes-instance of $\Lambda$ is tight.
    Hence $(g,f)$ is a tight solution-embedding reduction from $\Lambda_0$ to $\Lambda$.

    For the backward direction, suppose that $\Lambda$ is $\LOP$-complete, and let $\Pi_0 \in \NPS$ be arbitrary.
    Let $\Lambda_0$ be the zero-weight $\LOP$ problem derived from $\Pi_0$.
    Since $\Lambda$ is $\LOP$-complete, there is a tight solution-embedding reduction $(g,f)$ from $\Lambda_0$ to $\Lambda$.
    For an instance $I_0$ of $\Lambda_0$, let $g_0(I_0)$ denote the underlying instance $g(I_0)$ of $\Pi$;
    since $\Lambda$ has zero weights and threshold $0$, we have $\sol_t(g(I_0)) = \sol(g_0(I_0))$.
    Every yes-instance $I_0$ of $\Lambda_0$ is tight, so $f_{I_0}$ is solution-embedding for all yes-instances.
    If $I_0$ is a no-instance, then $\sol(g_0(I_0)) = \sol_t(g(I_0)) = \emptyset$ by the first condition of \Cref{def:tight-se-reduction}, and the injective map $f_{I_0}$ is trivially solution-embedding from $\emptyset$ to $\emptyset$.
    Thus, $g_0$ together with $f$ is an SE reduction from $\Pi_0$ to $\Pi$, and $\Pi$ is $\NPS$-complete.
\end{proof}

Intuitively, a zero-weight $\LOP$-complete problem acts as a reset mechanism for reduction chains.
Once an instance is mapped to a zero-weight problem, every feasible solution is optimal, so the optimization requirement disappears without losing the solution structure.
A subsequent TSE reduction can then reintroduce a nontrivial objective while preserving the property that the designated threshold captures exactly the optimal solutions.

\begin{lemma}
\label{lem:LOP-complete-derived-NPS}
    If $\Lambda$ is $\LOP$-complete, then the $\NPS$ problem $\Pi$ derived from $\Lambda$ by \Cref{lem:lop-to-nps} is $\NPS$-complete.
\end{lemma}
\begin{proof}
    Let $\Pi_0 \in \NPS$ be arbitrary, and let $\Lambda_0$ be the zero-weight $\LOP$ problem derived from $\Pi_0$.
    Since $\Lambda$ is $\LOP$-complete, there is a tight solution-embedding reduction $(g,f)$ from $\Lambda_0$ to $\Lambda$.
    Recall that $\sol_{\Pi}(g(I_0)) = \sol_t(g(I_0))$.
    Every yes-instance $I_0$ of $\Lambda_0$ is tight, so $f_{I_0}$ is solution-embedding for all yes-instances.
    If $I_0$ is a no-instance, then $\sol_{\Pi}(g(I_0)) = \sol_t(g(I_0)) = \emptyset$ by the first condition of \Cref{def:tight-se-reduction}, and the injective map $f_{I_0}$ is trivially solution-embedding.
    Hence $(g,f)$ is an SE reduction from $\Pi_0$ to $\Pi$.
\end{proof}

If one restricts the admissible weight functions, the converse of \Cref{lem:LOP-complete-derived-NPS} need not hold.
The reason is that $\LOP$-completeness requires the target of a reduction to be tight, and tightness constrains which threshold-feasible sets can occur.

\begin{remark}
\label{rem:converse-fails}
    Let $\Lambda_{{pow}}$ be \textsc{Satisfiability (Literal Universe)} with sense $\min$, arbitrary thresholds $t \in \Z$, and the weights restricted to powers of two: for an instance with literals $u_1, \ldots, u_{2n}$ in their fixed order, $w(u_k) = 2^{k-1}$.
    Then the $\NPS$ problem derived from $\Lambda_{{pow}}$ is $\NPS$-complete, but $\Lambda_{{pow}}$ is not $\LOP$-complete.
\end{remark}
\begin{proof}
    The weights of an instance have $O(n)$ bits each, so $\Lambda_{{pow}} \in \LOP$.
    For the first claim, map a formula $\varphi$ to the instance of $\Lambda_{{pow}}$ with threshold $t = 2^{2n} > w(\U)$, and use the identity on the literals.
    Then $\sol_t$ is exactly the set of satisfying assignments, so this is an SE reduction from \textsc{Sat-L} to the derived $\NPS$ problem, which is therefore $\NPS$-complete.

    For the second claim, note that distinct subsets of the literals have distinct weights.
    Hence every yes-instance of $\Lambda_{{pow}}$ has exactly one optimal solution, and a tight instance has exactly one threshold-feasible solution by \Cref{lem:tight-characterization}.
    Suppose there were a tight solution-embedding reduction $(g,f)$ to $\Lambda_{{pow}}$ from the zero-weight version of \textsc{Sat-L}, which is in $\LOP$.
    Let $\varphi$ be a formula with two distinct satisfying assignments $S_1 \neq S_2$; as a zero-weight instance, $\varphi$ is tight.
    Then $g(\varphi)$ is tight, and by \Cref{lem:se-set-equality}, the two distinct sets $f_\varphi(S_1) \neq f_\varphi(S_2)$ are of the form $S' \cap \U'_{emb}$ for threshold-feasible solutions $S'$ of $g(\varphi)$.
    This requires two distinct threshold-feasible solutions, a contradiction.
\end{proof}

The counterexample relies on restricting the admissible weight functions.
We do not know whether the converse holds for $\LOP$ problems that admit all non-negative integral weight functions.
However, restricting the weights is not harmful in itself:
by \Cref{lem:weight-monotonicity}, completeness for a small weight class implies completeness for every larger one, and 1-LOP contains many complete problems (cf. \Cref{tab:LOP-complete-compendium}).
What fails for $\Lambda_{pow}$ is that its weights separate all solutions, so every tight instance has a unique optimal solution.

\paragraph*{Satisfiability as a Linear Objective Problem}

In the realm of linear objective problems, \textsc{Satisfiability} introduces some interesting technicalities that we would like to discuss here.
As in the $\NPS$ setting, \textsc{Satisfiability} can be represented in several ways, depending on the choice of universe.
The corresponding $\LOP$ formulation equips every literal with a non-negative integral weight.

\begin{description}
    \item[\textsc{Satisfiability (Literal Universe)} as $\LOP$ problem]\hfill\\
    \textbf{Instances $\I$:} \quad Literals $L = \fromto{\ell_1, \overline \ell_1}{\ell_n, \overline \ell_n}$, Clauses $C \subseteq 2^L$.\\
    \textbf{Universe $\U$:} \quad Literal set $L = \fromto{\ell_1, \overline \ell_1}{\ell_n, \overline \ell_n}$.\\
    \textbf{Solutions $\sol$:} \quad $S \subseteq L$ such that $|S \cap \{\ell_i,\overline \ell_i\}| = 1$ for all $i$ and $|c \cap S| \geq 1$ for all $c \in C$.\\
    \textbf{Weight Function $w$:} \quad $w: L \to \Z_{\geq 0}$.\\
    \textbf{Threshold $t$:} \quad $t \in \Z$.
\end{description}

We now show that, indeed, \textsc{Satisfiability} is also the canonical $\LOP$-complete problem under tight solution-embedding reductions.

\begin{theorem}[Cook--Levin Theorem Adapted to $\LOP$]\label{thm:cook-levin-lop}
    \textsc{Satisfiability} (with literal or variable universe) is $\LOP$-complete with respect to polynomial-time tight solution-embedding reductions, even if $w: \U(I) \to \{0\}$ and $t = 0$.
\end{theorem}
\begin{proof}
    This follows from the Cook--Levin theorem adapted to $\NPS$ (\cite{DBLP:journals/corr/abs-2602-12350}, Theorem 7) and \Cref{lem:NP-SandLOP-completeness}.
\end{proof}

\textsc{Satisfiability} thus becomes the foundational problem for proving that problems in $\LOP$ are $\LOP$-complete, just as it was the foundation for proving problems to be $\NP$-complete.
For example, suppose that there is a tight solution-embedding reduction from \textsc{Sat} to an $\LOP$ problem $\Lambda$.
Let $\Lambda'$ be another $\LOP$ problem.
By \Cref{thm:cook-levin-lop}, $\Lambda' \leqTSE \textsc{Sat}$.
By assumption, $\textsc{Sat} \leqTSE \Lambda$.
So, by transitivity, $\Lambda' \leqTSE \Lambda$.
Therefore, $\Lambda$ is $\LOP$-complete.

\subsubsection{Subclasses of LOP}

The framework introduced above allows us to restrict the class of linear objective problems according to the form of their objective functions.
In this subsection, we consider two particularly useful subclasses.
The first is obtained by assigning the weight zero to every universe element, in which case the optimization component becomes trivial.
The second restricts the objective to unit weights and therefore corresponds to optimization by cardinality.

These subclasses are useful for two different reasons.
The zero-weight subclass provides the connection between $\LOP$ and $\NPS$, while the unit-weight subclass captures a broad class of classical combinatorial optimization problems.

\paragraph*{Zero-Weight Linear Objective Problems}

We first consider the subclass of zero-weight linear objective problems, $0\text{-}\LOP$, in which the weight functions are restricted to $w: \U(I) \rightarrow \{0\}$.

\begin{definition}[Zero-Weight Linear Objective Problems]
We define $0\text{-}\LOP$ to be the subclass of $\LOP$ consisting of problems whose weight functions satisfy
$$
    w^{(I)}(u)=0 \text{ for every instance $I \in \I$ and every $u \in \U(I)$}.
$$
\end{definition}

Consequently, whenever an instance is feasible, every feasible solution is optimal.
Thus, the objective function does not distinguish between feasible solutions.
For a minimization problem with threshold $t^{(I)}=0$, we have
$$
    \sol_t(I) = \sol(I) = \sol_{opt}(I).
$$
The analogous statement holds for maximization problems.
Hence, zero-weight problems provide a direct way of viewing an $\NPS$ problem as an $\LOP$ problem without changing its feasible or optimal solution structure.

In fact, \Cref{lem:NP-SandLOP-completeness} shows that an $\NPS$-complete problem equipped with the zero weight function and threshold $0$ is $\LOP$-hard, which is stronger than hardness for $0\text{-}\LOP$.
Since the containment is obvious, we directly obtain $\LOP$- and $0\text{-}\LOP$-completeness.
Conversely, every problem in $0\text{-}\LOP$ can be viewed as an $\NPS$ problem by incorporating the objective and threshold into its solution verifier.
Hence, the $\NPS$ problem derived from a $0\text{-}\LOP$ problem with $t=0$ is identical to the underlying $\NPS$ problem from which the $0\text{-}\LOP$ problem was derived.
As shown in \Cref{thm:cook-levin-lop}, \textsc{Satisfiability}, equipped with the zero weight function and threshold $0$, is $\LOP$-complete. 

\paragraph*{Cardinality Linear Objective Problems}

A different and particularly natural restriction is obtained by requiring every universe element to have weight one.
We call the resulting class \emph{cardinality linear objective problems}, denoted by $\CLOP$, where weight functions are of the form $w:\U(I)\rightarrow\{1\}.$
Thus, minimization problems seek feasible solutions of minimum cardinality, whereas maximization problems seek feasible solutions of maximum cardinality.

\begin{definition}[Cardinality Linear Objective Problems]
    We define $\CLOP$ to be the subclass of $\LOP$ consisting of problems whose weight functions satisfy
    $$
        w^{(I)}(u)=1 \text{ for every instance $I \in \I$ and every $u \in \U(I)$}.
    $$
\end{definition}

Many classical combinatorial optimization problems belong to $\CLOP$, including \textsc{Minimum Cardinality Vertex Cover}, \textsc{Minimum Cardinality Set Cover}, \textsc{Maximum Cardinality Independent Set}, and \textsc{Maximum Cardinality Set Packing}.

However, the restriction to unit weights does not prevent us from obtaining $\LOP$-complete problems.
In particular, we can again use \textsc{Satisfiability}, but now the choice of its solution universe becomes important.
Consider \textsc{Satisfiability (Literal Universe)}, in which the universe consists of the $2n$ literals
$$
    \U = \{\ell_1,\ldots,\ell_n\} \cup \{\overline{\ell}_1,\ldots,\overline{\ell}_n\}.
$$
Every feasible solution contains exactly one literal from each complementary pair and therefore has cardinality $n = |\U|/2.$
Consequently, assigning weight one to every literal gives every feasible solution the same objective value.
Setting the threshold to $t=|\U|/2$ therefore makes every feasible solution threshold-feasible and optimal, i.e. $\sol_t(I) = \sol(I) = \sol_{opt}(I)$ for every satisfiable instance $I$.
Hence, although the resulting problem belongs to the cardinality subclass $\CLOP$, its objective function does not distinguish between satisfying assignments.
The fact that this formulation is nevertheless complete for the entire class $\LOP$ follows from the solution-embedding version of the Cook--Levin construction.

\begin{theorem}[Cook--Levin Theorem Adapted to Unit-Weight Satisfiability]\label{thm:cook-levin-clop}
    \textsc{Satisfiability (Literal Universe)} with $w: \U(I) \to \{1\}$ and $t = |\U(I)|/2$ is $\LOP$-complete with respect to polynomial-time tight solution-embedding reductions.
\end{theorem}
\begin{proof}
    Let $\Lambda \in \LOP$.
    The Cook--Levin construction can be applied directly to the verifier for the threshold-feasible solutions $\sol_t$ of $\Lambda$.
    It produces an instance of \textsc{Satisfiability (Literal Universe)} together with a solution-embedding mapping from the threshold-feasible solutions $\sol_t$ of $\Lambda$ to the satisfying assignments of the constructed instance. 
    Then, we equip the \textsc{Satisfiability} instance with unit weights.
    Every satisfying assignment contains exactly one literal from each complementary pair and therefore has objective value exactly $|\U|/2$, which implies $\sol_{t-1}(I) = \emptyset$.
    Thus, setting $t = |\U|/2$ makes the threshold tight for every satisfiable instance.
    Consequently, the Cook--Levin solution-embedding reduction satisfies the tightness condition and is a tight solution-embedding reduction.
\end{proof}

Since \textsc{Satisfiability (Literal Universe)} is $\LOP$-hard and the weights are restricted to $w: \U(I) \to \{1\}$, it is the canonical complete problem for $\CLOP$.

\section{Min-Max Regret Problems with Interval Uncertainty}
\label{sec:regret-def}

In this section, we describe how to derive a min-max regret robust optimization problem with interval uncertainty from a linear objective optimization problem.

Let $\Lambda = (\I,\U,\sol,w,\sigma)$ be an $\LOP$ optimization problem, let $I \in \I$ be an instance, and let $c : \U(I) \to \R$ be a cost function.
The definition of regret depends on the optimization sense.
If $\sigma = \min$, the \emph{regret} of a solution $S \in \sol(I)$ with respect to $c$ is
$$
    \reg(S,c) := c(S) - \min\{c(S') : S' \in \sol(I)\}.
$$
If $\sigma = \max$, the regret of $S \in \sol(I)$ with respect to $c$ is
$$
    \reg(S,c) := \max\{c(S') : S' \in \sol(I)\} - c(S).
$$
In both cases, $\reg(S,c) = 0$ if and only if $S$ is optimal with respect to~$c$.

In an instance of a min-max regret problem, the costs (or weights) of the elements of $\U(I)$ are uncertain and are contained in an uncertainty set.
Given integers $\underline c_u, \overline c_u \in \Z$ with $\underline c_u \leq \overline c_u$ for every universe element $u \in \U(I)$, the \emph{interval uncertainty set} associated with the coefficient sequence $(\underline c_u, \overline c_u)_{u \in \U(I)}$ is
$$
    C := \set{c \mid c : \U(I) \to \R \text{ such that } c(u) \in [\underline c_u, \overline c_u] \ \forall u \in \U(I)}.
$$

\begin{definition}[Min-Max Regret Problem with Interval Uncertainty]
\label{def:interval-regret:opt}
    Let $\Lambda = (\I, \U, \sol, w, \sigma)$ be an $\LOP$ optimization problem.
    The min-max regret problem with interval uncertainty associated with $\Lambda$ is denoted by \textsc{Interval Min-Max Regret-$\Lambda$}.
    Its input is an instance $I \in \I$ together with integers $\underline c_u \leq \overline c_u$ for all $u \in \U(I)$, and the goal is to compute
    $$
        \min_{S \in \sol(I)} \ \max_{c \in C} \ \reg(S,c).
    $$
\end{definition}

We also consider the decision problem associated with the min-max regret problem.
In the decision version, the input additionally contains a threshold $t_R \in \Z$, and the question is whether
$$
    \min_{S \in \sol(I)} \ \max_{c \in C} \ \reg(S,c) \leq t_R.
$$
We use the convention $\min \emptyset = +\infty$.
Hence an instance with $\sol(I) = \emptyset$ is a no-instance of the decision version.
Note that \textsc{Interval Min-Max Regret-$\Lambda$} for $\Lambda = (\I, \U, \sol, w, \sigma)$ does not depend on the weight function $w$.

\subparagraph*{Min-max regret for $\LOP$ decision problems.}
$\LOP$-completeness is a property of decision problems, whereas \Cref{def:interval-regret:opt} is based on an optimization problem.
We connect the two by the following convention.
For an $\LOP$ decision problem $\Lambda = (\I, \U, \sol, w, t, \sigma)$, we define \textsc{Interval Min-Max Regret-$\Lambda$} to be \textsc{Interval Min-Max Regret-$\Lambda_{\textsc{opt}}$}, where $\Lambda_{\textsc{opt}}$ is the underlying $\LOP$ optimization problem of \Cref{def:LOP}.
That is, an instance $I = (I_0, w, t)$ of $\Lambda$ enters the min-max regret problem as the optimization instance $(I_0, w)$; its threshold $t$ is discarded.
Since the universe, the solution set and the weights of $I$ are those of $(I_0, w)$, we may and will write $I$ for both.

Let $r_\text{max}(S) := \max_{c \in C} \reg(S,c)$ denote the \emph{maximum regret} of $S \in \sol(I)$.
It is well known that the maximum regret is attained by a simple extreme scenario \cite{DBLP:journals/disopt/DeinekoW10}.
For $S \subseteq \U(I)$, the \emph{canonical cost scenario} $c_S : \U(I) \to \Z$ is defined by
$$
    c_S(u) :=   \begin{cases} \overline c_u & \text{if $u \in S$} \\
                    \underline c_u & \text{if $u \notin S$}
                \end{cases}
    \quad \text{if } \sigma = \min,
    \qquad
    c_S(u) :=   \begin{cases} \underline c_u & \text{if $u \in S$} \\
                    \overline c_u & \text{if $u \notin S$}
                \end{cases}
    \quad \text{if } \sigma = \max.
$$

\begin{lemma}
\label{lem:canonical-scenario}
    For every $S \in \sol(I)$, $r_\text{max}(S) = \reg(S, c_S)$.
    In particular, if $\sigma = \min$, then $r_\text{max}(S) = \max_{S' \in \sol(I)} \bigl(c_S(S) - c_S(S')\bigr)$, and if $\sigma = \max$, then $r_\text{max}(S) = \max_{S' \in \sol(I)} \bigl(c_S(S') - c_S(S)\bigr)$.
\end{lemma}
\begin{proof}
    Let $\sigma = \min$, and let $c \in C$ and $S' \in \sol(I)$. Then
    $$
        c(S) - c(S') = c(S \setminus S') - c(S' \setminus S) \leq c_S(S \setminus S') - c_S(S' \setminus S) = c_S(S) - c_S(S'),
    $$
    since $c_S$ takes upper bounds on $S$ and lower bounds outside $S$.
    Taking the maximum over $S'$ gives $\reg(S,c) \leq \reg(S,c_S)$, and $c_S \in C$ gives equality.
    For $\sigma = \max$, the same argument applies to $c(S') - c(S) = c(S' \setminus S) - c(S \setminus S')$.
\end{proof}

In particular, the maximum over the real-valued set $C$ is attained at an integral cost function.

With this observation, we can view an interval min-max regret problem as a two-player game between an $\exists$-player and an adversary, the $\forall$-player.
The $\exists$-player controls the outer minimum and chooses the solution $S$.
The adversary controls the maximum and chooses the cost function $c \in C$ and the solution $S'$; by \Cref{lem:canonical-scenario}, the choice of $c$ is determined by $S$.
With this perspective, it is easy to show containment in the class $\Sigma^p_2$.

\begin{lemma}
\label{lem:regret-containment-sigma-2}
    If $\Lambda = (\I, \U, \sol, w, \sigma)$ is an $\LOP$ optimization problem, then the decision version of \textsc{Interval Min-Max Regret-$\Lambda$} is contained in $\Sigma^p_2$.
\end{lemma}
\begin{proof}
    Let $J = (I, (\underline c_u, \overline c_u)_{u}, t_R)$ be an instance of \textsc{Interval Min-Max Regret-$\Lambda$}.
    By \Cref{def:interval-regret:opt} and \Cref{lem:canonical-scenario}, $J$ is a yes-instance if and only if
    $$
        \exists S \in \sol(I) \ \forall S' \in \sol(I) : \ \Delta(S,S') \leq t_R,
    $$
    where $\Delta(S,S') = c_S(S) - c_S(S')$ if $\sigma = \min$, and $\Delta(S,S') = c_S(S') - c_S(S)$ if $\sigma = \max$.
    Note that this also covers the case $\sol(I) = \emptyset$, in which $J$ is a no-instance.

    Since $|\U(I)| \leq \mathrm{poly}(|I|)$, we encode $S$ and $S'$ as characteristic vectors $y_1, y_2 \in \bin^{|\U(I)|}$.
    The verifier $V(J, y_1, y_2)$ accepts if and only if
    $$
        S \in \sol(I) \quad \text{and} \quad \bigl( S' \notin \sol(I) \ \text{ or } \ \Delta(S,S') \leq t_R \bigr).
    $$
    Membership in $\sol(I)$ can be decided in polynomial time since $\Lambda$ is based on an $\NPS$ problem, and $c_S$ and $\Delta(S,S')$ can be computed in polynomial time.
    Then $J$ is a yes-instance if and only if $\exists y_1 \, \forall y_2 : V(J,y_1,y_2) = 1$.
    Hence \textsc{Interval Min-Max Regret-$\Lambda$} is in $\Sigma^p_2$.
\end{proof}

\section{The Complexity of Min-Max Regret Satisfiability}
\label{sec:regret-sat}

We begin our complexity analysis by showing that a restricted version of \textsc{Interval Min-Max Regret-Sat} is $\Sigma^p_2$-complete.
In \Cref{sec:regret-min}, we use this problem as the starting point of a reduction showing that \textsc{Interval Min-Max Regret-$\Lambda$} is $\Sigma^p_2$-hard for every $\LOP$-complete problem $\Lambda$.

We regard \textsc{Satisfiability (Literal Universe)} as a minimization problem.
Its weight function $w$ plays no role here: by \Cref{def:interval-regret:opt}, the regret is computed with respect to the uncertain costs.
For \textsc{Satisfiability}, we denote these costs by $h$, to distinguish them from the costs $c$ of the target problem in \Cref{sec:regret-min}.
Following the notation of \Cref{sec:preliminaries}, the universe of a CNF formula $\varphi$ is its literal set $L$, and its solution set $\sol(\varphi)$ consists of the sets $L_\alpha$ for all satisfying assignments $\alpha$.

\begin{definition}[\textsc{Interval Min-Max Regret-Sat}]
\label{def:min-max-regret-sat}
    \textsc{Interval Min-Max Regret-Sat} is the problem of \Cref{def:interval-regret:opt} for \textsc{Satisfiability (Literal Universe)}.
    The input is a CNF formula $\varphi$ with literal set $L$ and clause set $C$, together with integers $\underline h_\ell \leq \overline h_\ell$ for all $\ell \in L$ and a threshold $t_R \in \Z$.
    Let $H$ be the interval uncertainty set defined by the $\underline h_\ell, \overline h_\ell$.
    The question is whether
    $$
        \min_{S \in \sol(\varphi)} \ \max_{h \in H} \ \left( h(S) - \min_{S' \in \sol(\varphi) } h(S') \right) \leq t_R.
    $$
    By the convention of \Cref{def:interval-regret:opt}, an instance with an unsatisfiable formula is a no-instance.
\end{definition}

The instances that we will construct in the following reduction for \textsc{Interval Min-Max Regret-Sat} will have a certain structure that is useful in the reduction in \Cref{sec:regret-min}.
We, therefore, define this restricted version of interval min-max regret-satisfiability separately and call it \textsc{Restricted Interval Min-Max Regret-Sat}.
First, we would like to have instances whose formula is guaranteed to be satisfiable.
We call a CNF formula \emph{$z$-padded} if it has a designated variable $z$ whose positive literal occurs in every clause.
Every $z$-padded formula is satisfiable, since setting $z = 1$ satisfies all clauses. Whether a formula is $z$-padded can be checked in polynomial time.
Second, the uncertainty set is defined by restricting $\underline h_\ell, \overline h_\ell \in \bin$.

\begin{definition}[\textsc{Restricted Interval Min-Max Regret-Sat}]
\label{def:restricted-min-max-regret-sat}
    \textsc{Restricted Interval Min-Max Regret-Sat} is the restriction of \textsc{Interval Min-Max Regret-Sat} to instances in which $\varphi$ is $z$-padded and $\underline h_\ell, \overline h_\ell \in \bin$ for all $\ell \in L$.
\end{definition}

We reduce from the canonical $\Sigma_2^p$-complete problem \textsc{$\exists\forall$DNF-Sat}.
Its input is a DNF formula $\psi(X,Y)$ whose variables are partitioned into $X$ and $Y$, and the question is whether $\exists X \forall Y: \psi(X, Y)$ is true.
This problem was introduced under the name $B_2$ and shown to be $\Sigma^p_2$-complete by Stockmeyer \cite{DBLP:journals/tcs/Stockmeyer76} and Wrathall \cite{DBLP:journals/tcs/Wrathall76}.

\begin{theorem}
\label{lem:minmaxregretsat}
    \textsc{Restricted Interval Min-Max Regret-Sat} is $\Sigma^p_2$-complete.
\end{theorem}
\begin{proof}
Containment in $\Sigma^p_2$ follows from \Cref{lem:regret-containment-sigma-2}, since \textsc{Satisfiability (Literal Universe)} is an $\LOP$ problem and the restriction can be checked in polynomial time.
For hardness, we reduce from \textsc{$\exists\forall$DNF-Sat}.

\proofsubparagraph*{Construction.}
Let $\psi_R(X,Y)$ be an instance of \textsc{$\exists\forall$DNF-Sat}.
Let $\varphi_R := \neg \psi_R$.
By De Morgan's rules, $\varphi_R$ is a CNF formula, obtained by replacing each product term $p$ of $\psi_R$ by the clause $\{\overline\ell : \ell \in p\}$.
Thus $\varphi_R(\alpha) = 1$ if and only if $\psi_R(\alpha) = 0$, and
$$
    \exists X \ \forall Y: \psi_R(X,Y) \ \text{ is true } \iff \exists X \ \neg\exists Y : \varphi_R(X,Y) \ \text{ is true}.
$$
We introduce a new variable $z$ and add the literal $z$ to every clause of $\varphi_R$.
The resulting CNF formula $\varphi$ is $z$-padded and $\varphi \equiv \varphi_R \lor z$ holds.
Let $L_X$, $L_Y$, and $L_z = \{z, \overline z\}$ be the literal sets of $X$, $Y$, and $z$, and let $L = L_X \cup L_Y \cup L_z$ be the literal set of $\varphi$.
We set
$$
    \underline h_\ell = 0,\ \overline h_\ell = 1 \ \text{ for } \ell \in L_X \cup L_z, \qquad
    \underline h_\ell = \overline h_\ell = 0 \ \text{ for } \ell \in L_Y,
$$
and $t_R := q := |X|$.
This instance is restricted, and it can be computed in polynomial time.

\proofsubparagraph*{The Regret.}
Let $S, S' \in \sol(\varphi)$.
By \Cref{lem:canonical-scenario}, the canonical scenario $h_S$ is $1$ on $S \cap (L_X \cup L_z)$ and $0$ elsewhere.
Since $S$ and $S'$ each contain exactly one literal of each of the $q+1$ complementary pairs in $L_X \cup L_z$, we obtain
$$
    h_S(S) - h_S(S') = |S \cap (L_X \cup L_z)| - |S \cap S' \cap (L_X \cup L_z)| = (q+1) - |S \cap S' \cap (L_X \cup L_z)|.
$$
Hence $r_\text{max}(S) \leq q+1$, and $r_\text{max}(S) = q+1$ if and only if there is $S' \in \sol(\varphi)$ that takes the opposite literal to $S$ on every variable of $X \cup \{z\}$.
Otherwise, $r_\text{max}(S) \leq q$.

\proofsubparagraph*{Correctness.}
We show that the min-max regret is at most $q$ if and only if $\exists X \neg\exists Y : \varphi_R(X,Y)$ is true.
Intuitively, the $\exists$-player selects $S$, and the adversary tries to answer with a satisfying assignment that is opposite to $S$ on $X \cup \{z\}$.
The $\exists$-player blocks the adversary by choosing her assignment on $X$ appropriately.

For the forward direction, suppose first that $\exists X \neg \exists Y \varphi_R(X,Y)$ is true.
Let $\alpha : X \to \bin$ be a truth assignment such that $\neg \exists Y \varphi_R(\alpha,Y)$ is true.
Let $S$ be the solution corresponding to an assignment $\overline \alpha$ that is opposite to $\alpha$, sets $z = 1$, and is arbitrary on $Y$.
Then $S \in \sol(\varphi)$, since $z = 1$.
Any assignment that is opposite to $S$ on $X \cup \{z\}$ equals $\alpha$ and sets $z = 0$.
Such an assignment satisfies $\varphi$ only if $\varphi_R(\alpha, Y)$ is true, which is impossible.
Hence $r_\text{max}(S) \leq q$, and the min-max regret is at most $q$.

For the backward direction, we prove the contrapositive.
Assume that $\forall X \exists Y \varphi_R(X,Y)$ is true (which is the negation of $\exists X \neg \exists Y \varphi_R(X,Y)$).
Let $S \in \sol(\varphi)$ be arbitrary, and let $\alpha_X: X \to \bin$, $\alpha_z: \{z\} \to \bin$ be the corresponding assignments to $X$ and $z$, respectively.
Further, let $\overline \alpha_X$ and $\overline \alpha_z$ be the assignments that are opposite to $\alpha_X$ and $\alpha_z$, respectively.
Choose $\beta_Y$ with $\varphi_R(\overline \alpha_X, \beta_Y) = 1$ and let $S'$ correspond to the assignment $\overline \alpha_X$, $\beta_Y$, and $\overline \alpha_z$.
Then $S'$ satisfies $\varphi$, because $\varphi_R$ is satisfied, and $S'$ is opposite to $S$ on $X \cup \{z\}$.
Hence $r_\text{max}(S) = q+1$.
Since $S$ was arbitrary, the min-max regret is $q+1 > q$.
\end{proof}

Since every instance of \textsc{Restricted Interval Min-Max Regret-Sat} is an instance of \textsc{Interval Min-Max Regret-Sat}, the identity map is a reduction from the former to the latter. Together with \Cref{lem:regret-containment-sigma-2}, we obtain the following.

\begin{corollary}
\label{cor:minmaxregretsat}
    \textsc{Interval Min-Max Regret-Sat} is $\Sigma^p_2$-complete.
\end{corollary}

\begin{remark}
\label{rem:minmaxregretsat-strength}
    The reduction in the proof of \Cref{lem:minmaxregretsat} shows that hardness holds under strong restrictions.
    \begin{itemize}
        \item All lower bounds are $0$, all upper bounds lie in $\bin$, and only the literals of $X \cup \{z\}$ have uncertain costs; the literals of $Y$ have the fixed cost $0$.
        \item The min-max regret of every constructed instance is either $q$ or $q+1$.
        Indeed, $r_\text{max}(S) \leq q+1$ for every $S \in \sol(\varphi)$, as shown above.
        Conversely, for every $S \in \sol(\varphi)$, the assignment that is opposite to $S$ on $X$, sets $z = 1$, and is arbitrary on $Y$ satisfies $\varphi$ and agrees with $S$ on at most one variable of $X \cup \{z\}$; hence $r_\text{max}(S) \geq q$.
        \item If every product term of $\psi_R$ has at most three literals, then every clause of $\varphi$ has at most four literals.
        Since \textsc{$\exists\forall$DNF-Sat} remains $\Sigma^p_2$-complete for formulas in 3-DNF \cite{DBLP:journals/tcs/Wrathall76}, \textsc{Restricted Interval Min-Max Regret-Sat} is $\Sigma^p_2$-complete already for 4-CNF formulas.
    \end{itemize}
\end{remark}

\section{Complexity of Min-Max Regret based on Minimization Problems}
\label{sec:regret-min}

Having established the hardness of \textsc{Restricted Interval Min-Max Regret-Sat}, we are now ready to state and prove our meta-theorem.
Recall from \Cref{sec:regret-def} that $\LOP$-completeness is a property of the decision problem $\Lambda$, whereas \textsc{Interval Min-Max Regret-$\Lambda$} is defined through the underlying optimization problem and ignores the threshold.
The proof connects the two through tight instances, on which the threshold-feasible solutions are exactly the optimal ones.

\begin{theorem}
    \label{thm:min-max-regret-reduction}
    For all $\LOP$-complete minimization problems $\Lambda = (\I,\U,\sol,w,t,\min)$, \textsc{Interval Min-Max Regret-$\Lambda$} is $\Sigma^p_2$-complete.
\end{theorem}

The idea of the proof is as follows.
As noted in \Cref{sec:regret-def}, \textsc{Interval Min-Max Regret-$\Lambda$} does not depend on the weight function $w$ of $\Lambda$.
Nevertheless, the hypothesis of \Cref{thm:min-max-regret-reduction} is a property of $\Lambda$ together with its weights.
The reason is that $w$ serves as a selector in the proof.
We start from a tight solution-embedding reduction from \textsc{Satisfiability} to $\Lambda$, whose target instances encode the satisfying assignments in their \emph{optimal} solutions with respect to $w$, but in general not in their feasible solutions.
We then add the cost intervals of the \textsc{Satisfiability} instance to a large multiple $M \cdot w$ of the nominal weights.
Any solution that is not optimal for $w$ then incurs a regret that is too large, so both players are forced to choose solutions that are optimal for $w$.
Among these, the regret only depends on the embedded part of the solutions, and tightness ensures that this embedded part ranges exactly over the satisfying assignments.

\begin{proof}
Containment in $\Sigma^p_2$ follows from \Cref{lem:regret-containment-sigma-2}.
For hardness, we reduce from \textsc{Restricted Interval Min-Max Regret-Sat}, which is $\Sigma^p_2$-complete by \Cref{lem:minmaxregretsat}.

\proofsubparagraph*{The LOP-completeness reduction.}
Let $\Lambda = (\I,\U,\sol,w,t,\min)$ be an $\LOP$-complete minimization problem and let $\Lambda_{\textsc{Sat}} = (\I_{\textsc{Sat}},\U_{\textsc{Sat}},\sol_{\textsc{Sat}},0,0,\min)$ be the $\LOP$ problem \textsc{Satisfiability (Literal Universe)} with $w \equiv 0$ and $t = 0$.
Since $\Lambda$ is $\LOP$-complete, there is a polynomial-time tight solution-embedding reduction $(g,f)$ from $\Lambda_{\textsc{Sat}}$ to $\Lambda$ that maps satisfiable CNF formulas $\varphi$ to instances $g(\varphi) \in \I$ of $\Lambda$ such that
$$
    \{f(S): S\in \sol_{\textsc{Sat}}(\varphi)\} = \{S' \cap f(\U(\varphi)) : S' \in \sol_{opt}(g(\varphi))\}.
$$
Given a $z$-padded instance $\varphi$, we know that $\varphi$ is satisfiable.
Because $w \equiv 0$ and $t = 0$, $\sol_{\textsc{Sat}}(\varphi) = \sol_{\textsc{Sat},0}(\varphi) \neq \emptyset$ and $\sol_{\textsc{Sat},-1}(\varphi) = \emptyset$.
By \Cref{lem:tight-characterization}, we have $\sol_{\textsc{Sat}}(\varphi) = \sol_{\textsc{Sat},t}(\varphi) = \sol_{\textsc{Sat},opt}(\varphi)$ and $(\varphi,w,t)$ is tight.
Since $(g,f)$ is a tight solution-embedding reduction, $I := g(\varphi,0,0)$ is tight and $\sol_t(I) \neq \emptyset$, so $I$ is also a yes-instance.

\proofsubparagraph*{Construction.}
Let $(\varphi, (\underline h_\ell, \overline h_\ell)_{\ell\in L}, q)$ be an instance of \textsc{Restricted Interval Min-Max Regret-Sat}.
If $\varphi$ is not $z$-padded, then we map the instance to a trivial no-instance.
We define an instance $(I, (\underline c_e, \overline c_e)_{e \in \U(g(\varphi))}, t_R)$ of \textsc{Interval Min-Max Regret-$\Lambda$} in the following way.
Let $n := |\U_{\textsc{Sat}}(\varphi)|$ and $M := 2n+1$.
\begin{itemize}
    \item The instance is given by $I = g(\varphi)$.
    \item The numbers $\underline c_e$, $\overline c_e$ for $e \in \U(g(\varphi))$ are defined by a case distinction.
        If $e \in \U_{emb}$, let $\ell := f^{-1}(e)$ and
        $$
            \underline c_e := M\,w^{(I)}(e) + \underline h_\ell, \qquad \overline c_e := M\,w^{(I)}(e) + \overline h_\ell.
        $$
        If $e \in \U(I) \setminus \U_{emb}$, let $\underline c_e = \overline c_e := M\,w^{(I)}(e)$.
        Let $C$ be the resulting interval uncertainty set.
    \item The threshold $t_R$ is given by $t_R := q$.
\end{itemize}
Following the convention of \Cref{sec:regret-def}, the threshold $t^{(I)}$ of $I$ is discarded; only the universe, the solution set and the weights of $I$ enter the min-max regret instance.
The construction is computable in polynomial time, because $g$, $f$, and all numbers $\underline c_e$ and $\overline c_e$ are polynomial-time computable.
Note that the set $C$ contains all functions of the form $c: \U(I) \to \mathbb R, e \mapsto M w^{(I)}(e) + \hat h(e)$,
where $\hat h := h \circ f^{-1}$ on $\Uemb$ with $h \in H$ and $\hat h := 0$ on $\U(I) \setminus \Uemb$.
We can therefore map a cost function $h \in H$ to $c = M w + \hat h \in C$.
Since $f$ is injective, $h \mapsto c$ is a bijection from $H$ onto $C$.
Since $\underline h_\ell, \overline h_\ell \in \bin$, every $h \in H$ satisfies $0 \leq h(T) \leq |L| \leq n$ for all $T \subseteq L$.

\proofsubparagraph*{Correctness.}

We start by showing the following three intermediate claims.
Consider an instance $I$ from above and let $w^* = \min_{S^* \in \sol(I)} w^{(I)}(S^*)$.

\begin{claim}
    \label{claim:meta-reduction:minimizer}
    For every $c \in C$, $\argmin_{S \in \sol(I)} c(S) \subseteq \sol_{opt}(I)$.
\end{claim}
\begin{claimproof}
    Let $S \in \sol(I) \setminus \sol_{opt}(I)$ and let $S^* \in \sol_{opt}(I)$.
    Since $w$ is integral, $w^{(I)}(S) \geq w^* + 1$, and hence
    $$
        c(S) \geq M(w^* + 1) > M w^* + n \geq c(S^*).
    $$    
\end{claimproof}

\begin{claim}
    \label{claim:meta-reduction:lowerbound}
    For every $S \in \sol(I) \setminus \sol_{opt}(I)$, we have $r_\text{max}(S) \geq M - n > n$.
\end{claim}
\begin{claimproof}
    Let $c \in C$ and $S^* \in \sol_{opt}(I)$.
    As in \Cref{claim:meta-reduction:minimizer},
    $$
        \reg(S,c) \geq c(S) - c(S^*) \geq M(w^* + 1) - (M w^* + n) = M - n = 2n+1 - n > n.
    $$
\end{claimproof}

\begin{claim}
    \label{claim:meta-reduction:upperbound}
    For every $S \in \sol_{opt}(I)$, we have
    $$
        r_\text{max}(S) = \max_{h \in H} \Bigl( h(f^{-1}(S \cap \Uemb)) - \min_{T \in \sol_{\textsc{Sat}}(\varphi)} h(T) \Bigr) \leq n.
    $$
\end{claim}
\begin{claimproof}
    By \Cref{claim:meta-reduction:minimizer}, for every $c \in C$, we have $\min_{S \in \sol(I)} c(S) = \min_{S \in \sol_{opt}(I)} c(S)$.
    Since $\hat h(e) = 0$ for all $e \in \U(I) \setminus \Uemb$, we have $c(S') = M w^* + h(f_I^{-1}(S' \cap \Uemb))$ for every $c \in C$ and $S' \in \sol_{opt}(I)$.
    Hence
    \begin{align*}
        \reg(S,c)
        &= h(f_I^{-1}(S \cap \Uemb)) - \min_{S' \in \sol_{opt}(I)} h(f_I^{-1}(S' \cap \Uemb)) \\
        &= h(f_I^{-1}(S \cap \Uemb)) - \min_{T \in \sol_{\textsc{Sat}}(\varphi)} h(T),
    \end{align*}
    where the last equality uses the solution-embedding property on the tight instance $\varphi$.
    Since $h \mapsto c$ is bijective, taking the maximum over $h \in H$ gives the maximum over $c \in C$, and thus the claimed equality.
    The bound follows from $0 \leq h(T) \leq n$.
\end{claimproof}

Because $\sol_{opt}(I) \neq \emptyset$, \Cref{claim:meta-reduction:lowerbound,claim:meta-reduction:upperbound} show that the minimum of $r_\text{max}$ over $\sol(I)$ is attained in $\sol_{opt}(I)$.
By \Cref{claim:meta-reduction:upperbound}, $r_\text{max}(S)$ depends on $S \in \sol_{opt}(I)$ only through $f^{-1}(S \cap \Uemb)$.
Since $\{f^{-1}(S \cap \Uemb) : S \in \sol_{opt}(I)\} = \sol_{\textsc{Sat}}(\varphi)$,
\begin{align*}
    \min_{S \in \sol(I)} r_\text{max}(S)
    &= \min_{S \in \sol_{opt}(I)} \ \max_{h \in H} \Bigl( h(f^{-1}(S \cap \Uemb)) - \min_{T' \in \sol_{\textsc{Sat}}(\varphi)} h(T') \Bigr) \\
    &= \min_{T \in \sol_{\textsc{Sat}}(\varphi)} \ \max_{h \in H} \Bigl( h(T) - \min_{T' \in \sol_{\textsc{Sat}}(\varphi)} h(T') \Bigr).
\end{align*}
The left-hand side is the min-max regret of the instance of $\Lambda$, while the right-hand side is the min-max regret of the \textsc{Satisfiability} instance.

Since both instances have the same threshold $t_R$, the constructed instance is a yes-instance if and only if the original instance is.
Hence \textsc{Interval Min-Max Regret-$\Lambda$} is $\Sigma^p_2$-hard, and together with containment, it is $\Sigma^p_2$-complete.
\end{proof}

We remark that \Cref{claim:meta-reduction:minimizer,claim:meta-reduction:lowerbound} do not use tightness:
they only force both players to choose solutions that are optimal for $w$.
Tightness is needed to identify these optimal solutions with the satisfying assignments of $\varphi$ via $\{f^{-1}(S \cap \Uemb) : S \in \sol_{opt}(I)\} = \sol_{\textsc{Sat}}(\varphi)$.
Without it, the solution-embedding property is only guaranteed for threshold-feasible solutions, which may differ from the optimal ones.

\begin{remark}
\label{rem:regret-max-direct}
    The same construction also applies directly to $\LOP$-complete maximization problems, which gives an alternative proof of \Cref{thm:regret-max}.
    Let $\Lambda$ be an $\LOP$-complete maximization problem, and let $L$ be the set of literals of $\varphi$.
    Define $k := 1 - h$, whose bounds $\underline k_\ell = 1 - \overline h_\ell$ and $\overline k_\ell = 1 - \underline h_\ell$ again lie in $\bin$, and use $k \mapsto c = Mw + \hat k$ in the construction.
    The analogues of \Cref{claim:meta-reduction:minimizer,claim:meta-reduction:lowerbound} show that both players choose solutions maximizing $w$.
    For such a solution $S$, the regret becomes $\max_{T' \in \sol(\varphi)} k(T') - k(f^{-1}(S \cap \Uemb))$.
    Since every $T \in \sol_{\textsc{Sat}}(\varphi)$ contains exactly $\frac{|L|}{2}$ literals, $k(T) = \frac{|L|}{2} - h(T)$, and this regret equals $h(f^{-1}(S \cap \Uemb)) - \min_{T' \in \sol_{\textsc{Sat}}(\varphi)} h(T')$.
    The duality of \Cref{sec:regret-max} yields the stronger statement that \textsc{Interval Min-Max Regret-$\Lambda$} and \textsc{Interval Min-Max Regret-$\Lambda^D$} are the same problem for every $\Lambda \in \LOP$.
\end{remark}

\section{Complexity of Min-Max Regret based on Maximization Problems}
\label{sec:regret-max}

In Section~\ref{sec:regret-min}, we showed that \textsc{Interval Min-Max Regret}-$\Lambda$ is $\Sigma^p_2$-complete for every $\LOP$-complete minimization problem~$\Lambda$.
We now extend this result to maximization problems.

The standard way to turn a maximization problem into a minimization problem is to replace the objective $w$ by $-w$.
This, however, does not fit our framework.
The resulting problem has negative weights, so it is not a problem in $\LOP$, and neither the notion of $\LOP$-completeness nor Theorem~\ref{thm:min-max-regret-reduction} applies to it.
We therefore use a different transformation, based on complementation.
Consider an $\LOP$ problem $\Lambda = (\I,\U,\sol,w,t,\sigma)$.
For every instance $I \in \I$ and every $S \subseteq \U(I)$,
$$
    w^{(I)}(S) = w^{(I)}(\U(I)) - w^{(I)}(\U(I) \setminus S).
$$
Maximizing $w^{(I)}(S)$ over the feasible solutions $S \in \sol(I)$ is therefore the same as minimizing $w^{(I)}(T)$ over their complements $T = \U(I) \setminus S$.
The objective remains linear and the weights remain unchanged, in particular non-negative.
We formalize this observation as a notion of duality.

The argument consists of three steps.
First, duality preserves $\NPS$-completeness (Section~\ref{sec:dual-nps}) and $\LOP$-completeness (Section~\ref{sec:dual-lop}).
Second, duality preserves regret:
a solution of a maximization problem has exactly the same regret as its complement in the dual minimization problem, under every cost vector (Section~\ref{sec:dual-regret}).
Third, applying Theorem~\ref{thm:min-max-regret-reduction} to the dual problem yields the result for maximization problems.
All three steps rest on one elementary fact:
complementing the solution sets preserves solution embeddings (Lemma~\ref{lem:complement-embedding}).

\subsection{Duals of Problems in $\NPS$}
\label{sec:dual-nps}

We summarize the concept of duality within the class $\NPS$ as presented in \cite{DBLP:journals/corr/abs-2602-12350}.
We start with the definition of a dual in $\NPS$.
The goal of this subsection is to show that if a problem is $\NPS$-complete, so is its dual.
This result also follows from results in~\cite{DBLP:journals/corr/abs-2602-12350}.
We include a short proof below because the same argument transfers to $\LOP$ in Section~\ref{sec:dual-lop}.

\begin{definition}[Dual of a problem in $\NPS$]
\label{def:dual-nps}
    Let $\Pi = (\mathcal{I}, \mathcal{U}, \mathcal{S})$ be a problem in $\NPS$.
    The \emph{dual} of $\Pi$ is the problem $\Pi^D = (\mathcal{I}, \mathcal{U},\mathcal{S}^D)$,
    which has the same instances and the same universes as~$\Pi$ and whose solution sets are
    $$
      \mathcal{S}^D(I) := \mathcal{S}(I)^D
      = \{ S \subseteq \mathcal{U}(I) : (\mathcal{U}(I) \setminus S) \in \mathcal{S}(I) \}
      \qquad \text{for all } I \in \mathcal{I}.
    $$
\end{definition}

The map $S \mapsto \U(I) \setminus S$ is a bijection between $\sol(I)$ and $\sol^D(I)$, and it is an involution.
In particular, $\sol^D(I) \neq \emptyset$ if and only if $\sol(I) \neq \emptyset$, and $(\sol^D)^D(I) = \sol(I)$.

\begin{lemma}
\label{lem:dual-nps}
    For every $\Pi \in \NPS$, the dual $\Pi^D$ is in $\NPS$, and $(\Pi^D)^D = \Pi$.
\end{lemma}
\begin{proof}
    Let $V$ be a polynomial-time verifier for $\Pi$.
    Then $V^D(I, C) := V(I, \mathcal{U}(I) \setminus C)$ is a polynomial-time verifier for $\Pi^D$, since $\mathcal{U}(I)$ has polynomial size.
    The instances and universes of $\Pi^D$ are those of $\Pi$, so $\Pi^D \in \NPS$.
    The identity $(\Pi^D)^D = \Pi$ follows from $(\sol^D)^D(I) = \sol(I)$.
\end{proof}

\begin{example}
\label{ex:dual-vc}
Consider the problem \textsc{Feasible Vertex Cover}.
An instance is a graph $G = (V, E)$, the universe is $V$, and $\mathcal{S}(G)$ is the set of all vertex covers of $G$.
Now, observe that a set $S \subseteq V$ is a vertex cover if and only if $V \setminus S$ is an independent set.
Hence $\sol^D(G)$ is the set of all independent sets of $G$, and the dual of \textsc{Feasible Vertex Cover} is \textsc{Feasible Independent Set}.

Now consider the $\NPS$ problem in which an instance additionally contains an integer $k$ and $\mathcal{S}(G, k)$ consists of all vertex covers of size at most~$k$.
Its dual has as solutions all independent sets of size at least $|V|- k$.
The cardinality bound $k$ is thus transformed into the bound $|V| - k$, and the direction of the inequality is reversed.
This is exactly how thresholds will be transformed in Section~\ref{sec:dual-lop}.
\end{example}

The central technical observation is that dualization is compatible with solution embeddings.

\begin{lemma}
\label{lem:se-dual}\label{lem:complement-embedding}
    Consider a solution-embedding reduction $(g,f)$ from $\Pi_1 = (\I_1,\U_1,\sol_1)$ to $\Pi_2 = (\I_2,\U_2,\sol_2)$.
    Then $(g, f)$ is a solution-embedding reduction from $\Pi_1^D$ to $\Pi_2^D$.
\end{lemma}
\begin{proof}
    
    The instances of $\Pi_i^D$ are those of $\Pi_i$, so $g$ maps instances of $\Pi_1^D$ to instances of $\Pi_2^D$.
    Moreover,
    $$
        \mathcal{S}_1^D(I) \neq \emptyset
        \iff \mathcal{S}_1(I) \neq \emptyset
        \iff \mathcal{S}_2(g(I)) \neq \emptyset
        \iff \mathcal{S}_2^D(g(I)) \neq \emptyset.
    $$
    For each instance $I$, the map $f_I$ is injective, uniformly polynomial-time computable, and solution-embedding.

    We now show that $f_I: \U_1(I) \to \U_2(g(I))$ is also solution-embedding between $\Pi_1^D$ and $\Pi_2^D$.
    To stay concise, we will write $f$ instead of $f_I$ and $\Uemb$ instead of $\Uemb(g(I)) = f_I(\U_1(I))$.
    We use the following three elementary identities.
    Since $f$ is injective and maps $\U_1(I)$ onto $\Uemb$, we have
    \begin{align}
        f(\U_1(I) \setminus A) &= \Uemb \setminus f(A) && \text{for all } A \subseteq \U_1(I), \label{eq:ce1}\\
        f^{-1}(\Uemb \setminus B) &= \U_1(I) \setminus f^{-1}(B) && \text{for all } B \subseteq \Uemb, \label{eq:ce2}\\
        (\U_2(g(I)) \setminus S_2) \cap \Uemb &= \Uemb \setminus (S_2 \cap \Uemb) && \text{for all } S_2 \subseteq \U_2(g(I)). \label{eq:ce3}
    \end{align}

    \begin{description}
        \item[SE Property~2.]
        Let $T_1 \in \sol_1^D(I)$.
        Then $S_1 := \U_1(I) \setminus T_1 \in \sol_1(I)$.
        By property~2 for $\Pi_1$ and $\Pi_2$, there is $S_2 \in \sol_2(g(I))$ with $f(S_1) = S_2 \cap \Uemb$. 
        Let $T_2 := \U_2(g(I)) \setminus S_2 \in \sol_2^D(g(I))$.
        By \eqref{eq:ce3} and \eqref{eq:ce1},
        $$
            T_2 \cap \Uemb = \Uemb \setminus f(S_1) = f(\U_1 \setminus S_1) = f(T_1).
        $$
    
        \item[SE Property~3.]
        Let $T_2 \in \sol_2^D(g(I))$.
        Then $S_2 := \U_2(g(I)) \setminus T_2 \in \sol_2(g(I))$, and by property~3 for $\Pi_1$ and $\Pi_2$, we have $S_1 := f^{-1}(S_2 \cap \Uemb) \in \sol_1(I)$.
        By \eqref{eq:ce3} and \eqref{eq:ce2},
        $$
            f^{-1}(T_2 \cap \Uemb) = f^{-1}\bigl(\Uemb \setminus (S_2 \cap \Uemb)\bigr) = \U_1(I) \setminus S_1 \in \sol_1^D(I). \qedhere
        $$
    \end{description}
\end{proof}

\begin{theorem}
\label{thm:dual-nps-complete}
    Let $\Pi \in \NPS$.
    Then $\Pi$ is $\NPS$-complete if and only if $\Pi^D$ is $\NPS$-complete.
\end{theorem}

\begin{proof}
Suppose that $\Pi$ is $\NPS$-complete, and let $\Pi' \in \NPS$ be arbitrary.
By Lemma~\ref{lem:dual-nps}, $\Pi'^D \in \NPS$, and hence $\Pi'^D \leqSE \Pi$.
By \Cref{lem:se-dual}, $\Pi' = (\Pi'^D)^D \leqSE \Pi^D$.
Since $\Pi^D \in \NPS$, the problem $\Pi^D$ is $\NPS$-complete.
The converse follows by applying the same argument to $\Pi^D$ and using $(\Pi^D)^D = \Pi$.
\end{proof}

\subparagraph*{Satisfiability and Duality.}
The dual of \textsc{Satisfiability} has a simple description. For a CNF
formula $I$, let $\overline{I}$ denote the formula obtained by replacing every
literal in every clause by its complementary literal. We call $\overline{I}$
the \emph{complement} of~$I$.

\begin{lemma}
\label{lem:sat-dual}
Consider \textsc{Satisfiability} with the literal universe or with the variable
universe. For every CNF formula $I$, we have $\mathcal{S}^D(I) =
\mathcal{S}(\overline{I})$.
\end{lemma}

\begin{proof}
For an assignment $\alpha$, let $\overline{\alpha}$ denote the complementary assignment.
A literal is true under $\alpha$ if and only if its complementary literal is true under $\overline{\alpha}$.
Hence $\alpha$ satisfies $I$ if and only if $\overline{\alpha}$ satisfies $\overline{I}$.

For the literal universe $L$, the solutions are the sets $L_\alpha$ for
satisfying assignments $\alpha$, and $L \setminus L_\alpha =
L_{\overline{\alpha}}$. Thus $T \in \mathcal{S}^D(I)$ if and only if $T =
L_{\overline{\alpha}}$ for some $\alpha$ satisfying $I$, which holds if and
only if $T \in \mathcal{S}(\overline{I})$. For the variable universe $X$, a
set $S \subseteq X$ encodes the assignment that is $1$ exactly on $S$, and $X
\setminus S$ encodes the complementary assignment. The same argument applies.
\end{proof}

Thus the dual of \textsc{Satisfiability} is \textsc{Satisfiability} applied to
complemented formulas. By Theorem~\ref{thm:dual-nps-complete}, it is
$\NPS$-complete. This can also be seen directly: $I \mapsto \overline{I}$,
together with the identity on the universe, is a solution-embedding reduction
from \textsc{Satisfiability} to its dual.

\subsection{Duals of Problems in $\LOP$}
\label{sec:dual-lop}

We now lift duality to linear objective problems.
Consider an $\LOP$ decision problem $\Lambda = (\I,\U,\sol,w,t,\sigma)$.
Complementation reverses the order of objective values, since
$$
    w^{(I)}(\U(I) \setminus S) = w^{(I)}(\U(I)) - w^{(I)}(S).
$$
The dual therefore has the opposite optimization sense, and a threshold
$t^{(I)}$ becomes $w^{(I)}(\U(I)) - t^{(I)}$ for its dual.
For an optimization sense $\sigma$, let $\overline{\sigma}$ denote the
opposite sense; that is, $\overline{\min} = \max$ and $\overline{\max} = \min$.

\begin{definition}[Dual of an $\LOP$ problem]
\label{def:dual-lop}
    Let $\Lambda_{opt} = (\I, \U, \sol, w, \sigma)$ be an $\LOP$ optimization problem based on $\Pi \in \NPS$.
    Its \emph{dual} is $\Lambda_{opt}^D := (\I, \U, \sol^D, w, \overline{\sigma})$, which is based on $\Pi^D$.
    
    Let $\Lambda = (\I, \U, \sol, w, t, \sigma)$ be the corresponding decision problem.
    For each instance $I = (I_0, w, t)$, let
    $$
        \delta(I) := \bigl(I_0,\, w,\, w(\U(I)) - t\bigr).
    $$
    The \emph{dual} of $\Lambda$ is $\Lambda^D := (\delta(\I), \U, \sol^D, w, t^D, \overline{\sigma})$, based on $\Pi^D$, where $\delta(\mathcal{I}) := \{\delta(I) : I \in \mathcal{I}\}$.
    That is, the instance $\delta(I)$ has the same underlying instance $I_0$ and the same weights $w^{(I)}$. However, the threshold is $t^{D,{(\delta(I))}} := w^{(I)}(\U(I)) - t^{(I)}$.
\end{definition}

The dual threshold $t^{D,{(\delta(I))}} := w^{(I)}(\U(I)) - t^{(I)}$ is negative whenever $t^{(I)} > w^{(I)}(\U(I))$.
This is one reason why \Cref{def:LOP} allows negative thresholds.
The map $\delta$ is computable in polynomial time, because $w^{(I)}(\U(I))$ is a sum of polynomially many numbers encoded in~$I$.
Its inverse, $\delta^{-1}(I_0,w,t') = (I_0,w,w(\U(I_0))-t')$, is also computable in polynomial time.
Because the weights are unchanged, the dual of a problem in $0\text{-}\LOP$ is in $0\text{-}\LOP$, and the dual of a problem in $\CLOP$ is in $\CLOP$.

\begin{lemma}
\label{lem:dual-lop}
    Let $\Lambda = (\mathcal{I}, \mathcal{U}, \mathcal{S}, w, t, \sigma) \in
    \LOP$, and let $I \in \mathcal{I}$. Then the following holds.
    \begin{enumerate}
        \item $\Lambda^D \in \LOP$ and $(\Lambda^D)^D = \Lambda$.
        \item $\mathcal{S}^D_{t^D}(\delta(I)) = \mathcal{S}_t(I)^D$;
            that is, the threshold-feasible solutions of $\delta(I)$ are exactly the complements of the threshold-feasible solutions of~$I$.
        \item If $\mathcal{S}(I) \neq \emptyset$, then $\mathcal{S}^D_{\mathrm{opt}}(\delta(I)) = \mathcal{S}_{\mathrm{opt}}(I)^D$.
        \item The threshold $t$ is tight for $I$ if and only if $t^D$ is tight for $\delta(I)$.
    \end{enumerate}
\end{lemma}
\begin{proof}
    Write $\mathcal{U} = \mathcal{U}(I)$ and $w = w^{(I)}$.
    Every $T \subseteq \mathcal{U}$ satisfies $w(T) = w(\mathcal{U}) - w(\mathcal{U} \setminus T)$.
    \begin{description}
        \item[Statement 1.]
        By Lemma~\ref{lem:dual-nps}, $\Pi^D \in \NPS$.
        The weights are non-negative integers, and the threshold is an integer, so $\Lambda^D \in \LOP$.
        Applying the construction twice returns the solution sets $(\mathcal{S}^D)^D = \mathcal{S}$, the sense $\overline{\overline{\sigma}} = \sigma$, and the threshold $w(\mathcal{U}) - (w(\mathcal{U}) - t) = t$.
        
        \item[Statement 2.]
        Let $\sigma = \min$. Then
        $$
          T \in \mathcal{S}^D_{t^D}(\delta(I))
          \iff \mathcal{U} \setminus T \in \mathcal{S}(I) \text{ and } w(T) \ge w(\mathcal{U}) - t
          \iff \mathcal{U} \setminus T \in \mathcal{S}(I) \text{ and } w(\mathcal{U} \setminus T) \le t,
        $$
        which holds if and only if $\mathcal{U} \setminus T \in \mathcal{S}_t(I)$.
        The case $\sigma = \max$ is symmetric.
        
        \item[Statement 3.]
        The bijection $S \mapsto \mathcal{U} \setminus S$ from $\mathcal{S}(I)$ to $\mathcal{S}^D(I)$ transforms objective values by $w(S) \mapsto w(\mathcal{U}) - w(S)$, which reverses their order.
        Hence it maps the $\sigma$-optimal solutions of $I$ exactly onto the $\overline{\sigma}$-optimal solutions of $\delta(I)$.
        
        \item[Statement 4.]
        If $\mathcal{S}(I) = \emptyset$, then also $\mathcal{S}^D(I) = \emptyset$, so $\mathcal{S}_t(I) = \emptyset = \mathcal{S}^D_{t^D}(\delta(I))$, and by \Cref{def:tight} neither threshold is tight.
        Otherwise, let $\sigma = \min$ and $z^* := \min_{S \in \sol(I)} w(S)$; the case $\sigma = \max$ is symmetric.
        Since $S \mapsto \U \setminus S$ is a bijection from $\sol(I)$ to $\sol^D(I)$ with $w(\U \setminus S) = w(\U) - w(S)$, the optimal value of the maximization instance $\delta(I)$ is $w(\U) - z^*$.
        By \Cref{lem:tight-characterization}, $t$ is tight for $I$ if and only if $t = z^*$, and $t^D = w(\U) - t$ is tight for $\delta(I)$ if and only if $w(\U) - t = w(\U) - z^*$.
    \end{description}
\end{proof}

\begin{example}
\label{ex:dual-lop}
    The dual of \textsc{Minimum Weight Vertex Cover} is \textsc{Maximum Weight Independent Set}, with the same vertex weights.
    The question ``Is there a vertex cover of weight at most $t$?'' becomes ``Is there an independent set of weight at least $w(V) - t$?''
    
    As a second example, consider \textsc{Knapsack} with item sizes $s$, capacity $B$, values $w$, and threshold $t$. Here $\mathcal{S}$ consists of all item sets $S$ with $s(S) \le B$, and the question is whether some $S \in \mathcal{S}$ has $w(S) \ge t$.
    A set $T$ belongs to the dual solution set if and only if $s(T) \ge s(\mathcal{U}) - B$.
    The dual is therefore the \textsc{Minimum Knapsack} (covering) problem:
    select items of total size at least $s(\mathcal{U}) - B$ and total value at most $w(\mathcal{U}) - t$.
\end{example}

We next show that tight solution-embedding reductions are compatible with duality.
Recall that, by \Cref{def:tight-se-reduction}, $f_I$ is injective for every instance~$I$, and solution-embedding from the threshold-feasible solutions $\mathcal{S}_t(I)$ to the threshold-feasible solutions $\mathcal{S}'_{t'}(g(I))$ for every tight instance~$I$.

\begin{lemma}
\label{lem:tse-dual}
    Let $\Lambda_1, \Lambda_2 \in \LOP$, and let $\delta_1, \delta_2$ be the maps of Definition~\ref{def:dual-lop} for $\Lambda_1$ and $\Lambda_2$, respectively.
    If $(g, f)$ is a tight solution-embedding reduction from $\Lambda_1$ to $\Lambda_2$, then $(g^D, f^D)$ with
    $$
      g^D := \delta_2 \circ g \circ \delta_1^{-1} \qquad \text{and} \qquad f^D_{\delta_1(I)} := f_I
    $$
    is a tight solution-embedding reduction from $\Lambda_1^D$ to $\Lambda_2^D$.
\end{lemma}
\begin{proof}
    The maps $g^D$ and $f^D$ are polynomial-time computable, since $\delta_1^{-1}$, $g$, $\delta_2$, and $f$ are.
    Since $\delta_1$ is a bijection from the instances of $\Lambda_1$ onto those of $\Lambda_1^D$, every instance of $\Lambda_1^D$ has the form $\delta_1(I)$ for a unique instance $I$ of $\Lambda_1$, so $f^D$ is well defined.
    Fix an instance $I$ of $\Lambda_1$, and let $I' := g(I)$, so that $g^D(\delta_1(I)) = \delta_2(I')$.
    Moreover, $\mathcal{U}(\delta_1(I)) = \mathcal{U}(I)$ and $\mathcal{U}(\delta_2(I')) = \mathcal{U}(I')$, so $f_I$ maps the universe of $\delta_1(I)$ into the universe of $\delta_2(I')$.
    This map is injective for every $I$, because $(g,f)$ is a tight solution-embedding reduction.
    
    \proofsubparagraph{Equivalence of instances.}
    By \Cref{lem:dual-lop}(2) and the fact that complementation is a bijection,
    $$
      \mathcal{S}^D_{t^D}(\delta_1(I)) \neq \emptyset
      \iff \mathcal{S}_t(I) \neq \emptyset
      \iff \mathcal{S}'_{t'}(I') \neq \emptyset
      \iff \mathcal{S}'^D_{t'^D}(\delta_2(I')) \neq \emptyset.
    $$
    
    \proofsubparagraph{Tightness.}
    Suppose that $t^D$ is tight for $\delta_1(I)$.
    By \Cref{lem:dual-lop}(4), $t$ is tight for $I$.
    Hence $t'$ is tight for $I'$, and by \Cref{lem:dual-lop}(4) again, $t'^D$ is tight for $\delta_2(I')$.
    
    \proofsubparagraph{Solution embedding.}
    Suppose again that $t^D$ is tight for $\delta_1(I)$.
    Then $I$ is tight, as shown above, so $f_I$ is solution-embedding from $\mathcal{S}_t(I)$ to $\mathcal{S}'_{t'}(I')$.
    By the argument of \Cref{lem:complement-embedding}, it is solution-embedding from $\mathcal{S}_t(I)^D$ to $\mathcal{S}'_{t'}(I')^D$.
    By \Cref{lem:dual-lop}(2), these are exactly $\mathcal{S}^D_{t^D}(\delta_1(I))$ and $\mathcal{S}'^D_{t'^D}(\delta_2(I'))$.
\end{proof}

\begin{theorem}
\label{thm:dual-lop-complete}
    Let $\Lambda \in \LOP$.
    Then $\Lambda$ is $\LOP$-complete if and only if $\Lambda^D$ is $\LOP$-complete.
\end{theorem}
\begin{proof}
    Suppose that $\Lambda$ is $\LOP$-complete, and let $\Lambda' \in \LOP$ be arbitrary.
    By \Cref{lem:dual-lop}(1), $\Lambda'^D \in \LOP$, and hence $\Lambda'^D \leqTSE \Lambda$.
    By \Cref{lem:tse-dual} and \Cref{lem:dual-lop}(1), $\Lambda' = (\Lambda'^D)^D \leqTSE \Lambda^D$.
    Thus $\Lambda^D$ is $\LOP$-complete.
    The converse follows by applying the same argument to $\Lambda^D$.
\end{proof}

\begin{corollary}
\label{cor:max-to-min}
    For every $\LOP$-complete maximization problem $\Lambda$, the dual $\Lambda^D$ is an $\LOP$-complete minimization problem with the same weights.
\end{corollary}

For example, \Cref{thm:dual-lop-complete} and \Cref{ex:dual-lop} show that \textsc{Maximum Weight Independent Set} is $\LOP$-complete, because \textsc{Minimum Weight Vertex Cover} is $\LOP$-complete (\Cref{sec:compendium}).

\subsection{Min-Max Regret for Maximization Problems}
\label{sec:dual-regret}

The regret of a solution depends only on the feasible solution set, the optimization sense, and the realized cost vector.
It does not depend on the weight function $w$.
Duality preserves regret exactly.

\begin{lemma}[Duality preserves regret]
\label{lem:regret-dual}
    Let $\Lambda_{opt} = (\I, \U, \sol, w, \sigma)$ be an $\LOP$ optimization problem,
    let $I \in \I$ with $\sol(I) \neq \emptyset$, and let $c: \U(I) \to \R$ be an arbitrary cost function.
    For every $S \in \sol(I)$,
    $$
      \reg_{\Lambda}(S, c) = \reg_{\Lambda^D}(\U(I) \setminus S, c).
    $$
\end{lemma}

\begin{proof}
    Write $\U = \U(I)$ and $T = \U \setminus S$.
    Let $\sigma = \max$, so that $\Lambda^D$ is a minimization problem.
    As $S'$ ranges over $\sol(I)$, the complement $T' = \U \setminus S'$ ranges over $\sol^D(I)$, and $c(S') = c(\U) - c(T')$.
    Hence
    \begin{align*}
        \reg_{\Lambda}(S, c) &= \max_{S' \in \sol(I)} c(S') - c(S) = \Bigl(c(\U) - \min_{T' \in \sol^D(I)} c(T')\Bigr) - \bigl(c(\U) - c(T)\bigr) \\
        &= c(T) - \min_{T' \in \sol^D(I)} c(T') = \reg_{\Lambda^D}(T, c).
    \end{align*}
    The case $\sigma = \min$ is symmetric.
\end{proof}

Because $S \mapsto \U(I) \setminus S$ is a bijection from $\sol(I)$ to $\sol^D(I)$, and because the interval uncertainty set $C$ is the same for both problems, Lemma~\ref{lem:regret-dual} implies
\begin{equation}
  \min_{S \in \sol(I)} \max_{c \in C} \reg_{\Lambda}(S, c)
  = \min_{T \in \sol^D(I)} \max_{c \in C} \reg_{\Lambda^D}(T, c).
  \label{eq:regret-dual}
\end{equation}
Since $\Lambda_{opt}$ and $\Lambda^D_{opt}$ have the same instance set, an instance $(I, (\underline c_u, \overline c_u)_{u}, t_R)$ of \textsc{Interval Min-Max Regret}-$\Lambda^D$ is also an instance of \textsc{Interval Min-Max Regret}-$\Lambda$.
By \eqref{eq:regret-dual}, it is a yes-instance of one problem if and only if it is a yes-instance of the other.
The two problems are therefore identical as languages.

We can now prove our main result for maximization problems.

\begin{theorem}
\label{thm:regret-max}
    For all $\LOP$-complete maximization problems $\Lambda = (\I, \U, \sol, w, t, \max)$, \textsc{Interval Min-Max Regret}-$\Lambda$ is $\Sigma^p_2$-complete.
\end{theorem}

\begin{proof}
    Containment in $\Sigma^p_2$ follows from Lemma~\ref{lem:regret-containment-sigma-2}, which covers both optimization senses.
    For hardness, note that $\Lambda^D$ is an $\LOP$-complete minimization problem by Corollary~\ref{cor:max-to-min}.
    By Theorem~\ref{thm:min-max-regret-reduction}, \textsc{Interval Min-Max Regret}-$\Lambda^D$ is $\Sigma^p_2$-hard.
    By \eqref{eq:regret-dual}, the identity map is a polynomial-time many-one reduction from \textsc{Interval Min-Max Regret}-$\Lambda^D$ to \textsc{Interval Min-Max Regret}-$\Lambda$.
    Hence \textsc{Interval Min-Max Regret}-$\Lambda$ is $\Sigma^p_2$-hard.
\end{proof}

Combining Theorems~\ref{thm:min-max-regret-reduction} and~\ref{thm:regret-max}, we obtain our main application result.

\begin{corollary}
\label{cor:regret-all}
    For every $\LOP$-complete problem $\Lambda$, whether a minimization or a maximization problem, \textsc{Interval Min-Max Regret}-$\Lambda$ is $\Sigma^p_2$-complete.
\end{corollary}

Since the definition of \textsc{Interval Min-Max Regret}-$\Lambda$ ignores $w$ and the reduction does not introduce any restrictions on $w$, its hardness is established as soon as $\Lambda$ is $\LOP$-complete for \emph{some} weight class.

In particular, since \textsc{Knapsack} with its natural objective is known to be $\LOP$-complete (Appendix~\ref{sec:compendium}), \Cref{cor:regret-all} recovers the $\Sigma^p_2$-completeness of interval min-max regret knapsack due to Deineko and Woeginger~\cite{DBLP:journals/disopt/DeinekoW10}.

\section{Conclusion}
\label{sec:conclusion}

We introduced linear objective problems ($\LOP$), a framework that extends $\NP$ with solutions ($\NPS$) by a linear objective over the universe elements, and we defined tight solution-embedding reductions together with the resulting notion of $\LOP$-completeness.
The crucial property of these reductions is that they map optimal solutions of the source problem onto optimal solutions of the target problem, even though recognizing an optimal solution is $\sf coNP$-hard in general.
In this sense, $\LOP$ is the appropriate framework when a level is characterized by solving an optimization problem, whereas $\NPS$ is appropriate when a level only selects a feasible solution.
We showed that \textsc{Satisfiability} is $\LOP$-complete, even with unit weights, and that $\LOP$-completeness is preserved under duality, which relates minimization and maximization problems.
The compendium in \Cref{sec:compendium} shows that many classical problems are $\LOP$-complete.

We applied the framework to min-max regret optimization under interval uncertainty.
We first showed that \textsc{Interval Min-Max Regret-Sat} is $\Sigma_2^p$-complete, even under strong restrictions.
We then used this problem as an anchor for a meta-reduction showing that the min-max regret version of every $\LOP$-complete minimization problem is $\Sigma_2^p$-complete.
Via duality, the same holds for maximization problems.
Hence, instead of proving completeness separately for each nominal problem, it suffices to show that the nominal problem is $\LOP$-complete.
This recovers the known results for \textsc{Interval Min-Max Regret-Knapsack}~\cite{DBLP:journals/disopt/DeinekoW10} and \textsc{Interval Min-Max Regret-Maximum Benefit Set Cover}~\cite{DBLP:journals/coap/CocoSN22} and yields completeness for the further problems in the compendium.

The $\LOP$ framework also places earlier work in a common setting.
The completeness results for blocking and interdiction problems~\cite{DBLP:journals/tcs/GruneW26}, Stackelberg pricing games~\cite{DBLP:conf/esa/GruneHRW26}, and recoverable robust optimization~\cite{DBLP:conf/mfcs/GruneW25} rely on the same underlying techniques.
They were obtained before the framework was developed in the generality presented here, and they are consistent with it.
We view $\LOP$ as a systematic account of the ideas behind these results.

\subparagraph*{Limitations.}
The framework has two restrictions.
First, the objective must be linear over the chosen universe.
Problems whose natural objective is not linear in this sense, such as \textsc{Uncapacitated Facility Location}, do not fit the framework in their natural form.
Second, $\LOP$-completeness is a strictly stronger requirement than $\NPS$-completeness of the derived problem once the admissible weights are restricted.
\Cref{rem:converse-fails} shows this for \textsc{Satisfiability} with weights that are powers of two:
every tight instance then has a unique optimal solution, so the problem is not $\LOP$-complete although its derived $\NPS$ problem is $\NPS$-complete.
Consequently, for problems whose feasibility version is easy (Type~B in \Cref{sec:compendium}), $\LOP$-completeness does not follow from known $\NPS$-completeness results, and each such problem needs an additional argument such as the Bound Lemma (\Cref{lem:bound}).

\subparagraph*{Open questions.}
The most prominent open question is whether the framework extends from two levels to $k > 2$ levels.
For bilevel problems, the lower level has a fixed linear objective.
With each additional level, the cost function seen by the follower changes, and a reduction would have to account for this.
We therefore do not know whether a meta-theorem of the form ``$\LOP$-complete implies $\Sigma_{k}^p$-complete'' can be obtained, and this is the reason we state our results for bilevel optimization.
A second question concerns the converse of \Cref{lem:LOP-complete-derived-NPS}:
if a problem admits all non-negative integral weight functions, does $\NPS$-completeness of the derived problem imply $\LOP$-completeness?
\Cref{rem:converse-fails} only rules this out for restricted weight classes.
Additionally, it would be desirable to have a characterization of the Type~B problems for which a Bound Lemma argument exists, instead of verifying this case by case.
Finally, it would also be interesting to extend the min-max regret results to other uncertainty sets, such as discrete scenarios.

\appendix
\section{Compendium of $\LOP$-Complete Problems}
\label{sec:compendium}

This section collects natural optimization problems that are $\LOP$-complete with respect to tight solution-embedding reductions.
It builds on the compendium of $\NPS$-complete problems in the appendix of \cite{DBLP:journals/corr/abs-2602-12350}.
We first introduce weight classes and two general tools, which reduce most entries to a one-line argument.
We then work out \textsc{Vertex Cover} in detail and summarize all problems in \Cref{tab:LOP-complete-compendium}, followed by a short certificate for each entry.

\subsection{Weight Classes}

Let $\Pi = (\I_0, \U, \sol)$ be an $\NPS$ problem, let $\sigma \in \{\min,\max\}$, and let $W \subseteq \Z_{\geq 0}$.
We denote by $\Lambda_W(\Pi, \sigma)$ the $\LOP$ problem based on $\Pi$ with sense $\sigma$ whose instances are all $(I_0, w, t)$ with $I_0 \in \I_0$, $w : \U(I_0) \to W$, and $t \in \Z$.
We write $0\text{-}\LOP$ and $\CLOP$ for the subclasses with $W = \{0\}$ and $W = \{1\}$, respectively, and we call $\Lambda_{\Z_{\geq 0}}(\Pi,\sigma)$ the version with \emph{general weights}.

\begin{lemma}[Monotonicity]
\label{lem:weight-monotonicity}
    Let $W \subseteq W' \subseteq \Z_{\geq 0}$. If $\Lambda_W(\Pi,\sigma)$ is $\LOP$-complete, then so is $\Lambda_{W'}(\Pi,\sigma)$.
\end{lemma}
\begin{proof}
    Every instance of $\Lambda_W(\Pi,\sigma)$ is an instance of $\Lambda_{W'}(\Pi,\sigma)$ with the same solution sets, threshold-feasible sets, and optimal solutions.
    Hence the identity, together with the identity on the universe, is a tight solution-embedding reduction from $\Lambda_W(\Pi,\sigma)$ to $\Lambda_{W'}(\Pi,\sigma)$.
    The claim follows from \Cref{lem:TSE-transitive}.
\end{proof}

Consequently, completeness with zero weights or with unit weights implies completeness with general weights, and the informative statement about a problem is the smallest weight class for which it is complete.
The classes $\{0\}$ and $\{1\}$ are not comparable, so neither implies the other.

The problems of this section fall into two types, which behave differently with respect to weights.
\begin{itemize}
    \item \emph{Type A: the feasibility problem $\Pi$ is $\NPS$-complete.}
    Examples are \textsc{Satisfiability}, \textsc{Hamiltonian Path}, and \textsc{Subset Sum}.
    By \Cref{lem:NP-SandLOP-completeness} (using $t \equiv 0$), $\Lambda_{\{0\}}(\Pi, \sigma)$ is $\LOP$-complete for both senses, so the objective plays no role for completeness.
    \item \emph{Type B: whether an instance of $\Pi$ has a feasible solution can be decided in polynomial time.}
    Examples are \textsc{Vertex Cover}, \textsc{Dominating Set}, and \textsc{Knapsack}, where a feasible solution always exists, and \textsc{Set Cover} and \textsc{Steiner Tree}, where feasibility is easy to check.
    Then $\Lambda_{\{0\}}(\Pi,\sigma)$ is decidable in polynomial time, so it is not $\LOP$-complete unless $\PTIME = \NP$.
    For these problems, the $\NPS$ compendium of \cite{DBLP:journals/corr/abs-2602-12350} treats the threshold version, which is the $\NPS$ problem derived from a weighted version by \Cref{lem:lop-to-nps}.
    Since $\NPS$-completeness of the derived problem does not imply $\LOP$-completeness (\Cref{rem:converse-fails}), an additional argument is needed.
\end{itemize}

\subsection{Two Tools}

The following lemma provides an additional argument for many entries.
It requires that the threshold of the constructed instance bounds the weight of \emph{every} solution $S \in \sol(I)$, and not only of the threshold-feasible ones $S \in \sol_t(I)$.
An SE reduction for the derived $\NPS$ problem controls the threshold-feasible set $\sol_t(I)$, but not which of its members are optimal.
If some threshold-feasible solution had weight strictly below $t^{(I)}$ for $\sigma = \min$, then $\sol_{t-1}(I) \neq \emptyset$ and $g(I_0)$ would not be tight.
The bound forces every threshold-feasible solution to have weight exactly $t^{(I)}$, so that $\sol_t(I) = \sol_{opt}(I)$.

\begin{lemma}[Bound Lemma]
\label{lem:bound}
    Let $\Lambda = (\I, \U, \sol, w, t, \sigma) \in \LOP$, and let $\Pi_\Lambda$ be the $\NPS$ problem derived from $\Lambda$ by \Cref{lem:lop-to-nps}.
    Let $\Pi_0$ be an $\NPS$-complete problem, and let $(g, f)$ be a solution-embedding reduction from $\Pi_0$ to $\Pi_\Lambda$.
    Suppose that for every yes-instance $I_0$ of $\Pi_0$, the instance $I = g(I_0)$ satisfies
    $
        w^{(I)}(S) \geq t^{(I)} \text{ for all } S \in \sol(I) \text{ if } \sigma = \min,
    $
    and
    $    
        w^{(I)}(S) \leq t^{(I)} \text{ for all } S \in \sol(I) \text{ if } \sigma = \max.
    $
    Then $\Lambda$ is $\LOP$-complete.
\end{lemma}
\begin{proof}
    Let $\Lambda_0 = (\I_0,\U_0,\sol_0,w_0,t_0,\sigma_0)$ be the zero-weight $\LOP$ problem derived from $\Pi_0$, with an arbitrary sense $\sigma_0$, which is $\LOP$-complete by \Cref{lem:NP-SandLOP-completeness}.
    We show that $(g,f)$ is a tight solution-embedding reduction from $\Lambda_0$ to $\Lambda = (\I,\U,\sol,w,t,\sigma)$, identifying an instance of $\Lambda_0$ with its underlying instance $I_0$.
    We have $\sol_{t,0}(I_0) = \sol_0(I_0)$ and, by \Cref{lem:lop-to-nps}, $\sol_{\Pi_\Lambda}(g(I_0)) = \sol_t(g(I_0))$.
    As $(g, f)$ is an SE reduction from $\Pi_0$ to $\Pi_\Lambda$, it follows that $\sol_{t,0}(I_0) \neq \emptyset$ if and only if $\sol_t(g(I_0)) \neq \emptyset$, so the first condition of \Cref{def:tight-se-reduction} holds.
    Moreover, since $(g,f)$ is an SE reduction, every $f_{I_0}$ is injective and computable uniformly in polynomial time, and it is solution-embedding from $\sol_{t,0}(I_0)$ to $\sol_t(g(I_0))$ for every instance $I_0$.
    
    It remains to show that $I := g(I_0)$ is tight whenever $I_0$ is tight.
    Let $\sigma = \min$, and let $I_0$ be tight.
    Then $I_0$ is a yes-instance, so $\sol_t(I) \neq \emptyset$ by the first condition, and $w^{(I)}(S) \geq t^{(I)}$ for all $S \in \sol(I)$ by assumption.
    Every $S \in \sol_t(I)$ additionally satisfies $w^{(I)}(S) \leq t^{(I)}$, so $t^{(I)}$ is a lower bound on the weight of every solution that is attained, i.e., the optimal value.
    Thus, $t^{(I)}$ is tight by \Cref{lem:tight-characterization}.
    The case $\sigma = \max$ is symmetric, and the claim follows from \Cref{lem:TSE-transitive}.
\end{proof}

The simplest application is to Type A problems whose solutions all have the same cardinality; this generalizes \Cref{thm:cook-levin-clop}.

\begin{corollary}[Constant Cardinality]
\label{cor:constant-cardinality}
    Let $\Pi$ be $\NPS$-complete, and suppose that there is a polynomial-time computable function $k$ such that $|S| = k(I_0)$ for every instance $I_0$ of $\Pi$ and every $S \in \sol(I_0)$.
    Then $\Lambda_{\{1\}}(\Pi, \sigma)$ is $\LOP$-complete for both senses.
\end{corollary}
\begin{proof}
    Apply \Cref{lem:bound} with $\Pi_0 = \Pi$, the map $g(I_0) := (I_0, w \equiv 1, k(I_0))$, and the identity on the universe.
    Every feasible solution of $g(I_0)$ has weight exactly $k(I_0)$, so for either sense its threshold-feasible solutions are all of $\sol(I_0)$, and $(g, \mathrm{id})$ is an SE reduction from $\Pi$ to the derived $\NPS$ problem.
    Since $k(I_0)$ is both a lower and an upper bound on the weight of every feasible solution, \Cref{lem:bound} applies for both senses.
\end{proof}

The second tool is duality.
By \Cref{thm:dual-lop-complete}, a problem is $\LOP$-complete if and only if its dual is.
Since duality does not change the weights, it preserves each weight class.
Thus every entry of Type~B yields a further complete problem with the opposite sense, for example \textsc{Maximum Independent Set} from \textsc{Minimum Vertex Cover} and \textsc{Minimum Knapsack} from \textsc{Knapsack}.

\paragraph*{A Worked Example: Vertex Cover}

\begin{proposition}
\label{prop:vc-lop-complete}
    \textsc{Minimum Cardinality Vertex Cover}, that is, $\Lambda_{\{1\}}(\Pi_{\mathrm{VC}}, \min)$ with $\Pi_{\mathrm{VC}}$ the $\NPS$ problem of all vertex covers with universe $V$, is $\LOP$-complete.
    Consequently, \textsc{Minimum Weight Vertex Cover} and \textsc{Maximum (Weight) Independent Set} are $\LOP$-complete.
\end{proposition}
\begin{proof}
    We use the classical reduction from \textsc{3-Satisfiability (Literal Universe)}, which is $\NPS$-complete \cite{DBLP:journals/corr/abs-2602-12350}.
    Let $\varphi$ have variables $x_1,\ldots,x_n$ and clauses $c_1,\ldots,c_m$.
    The graph $G$ contains, for each variable, the two literal vertices $x_i, \overline x_i$ joined by an edge, and for each clause $c_j = (\ell_{j1} \lor \ell_{j2} \lor \ell_{j3})$, a triangle on vertices $a_{j1}, a_{j2}, a_{j3}$ together with the edges $\{a_{jk}, \ell_{jk}\}$ for $k \in \{1,2,3\}$.
    All weights are $1$, the threshold is $t = n + 2m$, and $f$ maps each literal to its literal vertex.

    \proofsubparagraph{Bound.}
    The $n$ variable edges and the $m$ triangles are pairwise vertex-disjoint.
    A vertex cover contains at least one endpoint of each variable edge and at least two vertices of each triangle, so every vertex cover has at least $n + 2m$ vertices.

    \proofsubparagraph{Solution embedding.}
    A vertex cover $S$ of size $n+2m$ contains exactly one literal vertex per variable, which defines an assignment $\alpha$ with $S \cap f(L) = f(L_\alpha)$, and exactly two vertices of each triangle.
    The remaining vertex $a_{jk}$ of a triangle is not in $S$, so its literal $\ell_{jk}$ is in $S$, and clause $c_j$ is satisfied by $\alpha$.
    Conversely, for a satisfying assignment $\alpha$, the set $L_\alpha$ together with, for each clause, the two triangle vertices other than one vertex whose literal is true is a vertex cover of size $n+2m$.
    Hence $(g,f)$ is an SE reduction from \textsc{3-Satisfiability} to the threshold version of vertex cover.

    By \Cref{lem:bound}, the unit-weight problem is $\LOP$-complete. The remaining claims follow from \Cref{lem:weight-monotonicity} and from duality, since the dual of vertex cover is independent set (\Cref{ex:dual-lop}).
\end{proof}

\subsection{The Compendium}

\Cref{tab:LOP-complete-compendium} lists the problems together with the smallest weight class for which we establish completeness.
By \Cref{lem:weight-monotonicity}, completeness extends to every larger class, in particular to general weights.
For Type A problems, the sense is irrelevant for completeness, and we write ``any''.

\begin{table}[!ht]
\centering
\small
\setlength{\tabcolsep}{4pt}
    \begin{tabular}{lllll}
        \hline
        \textbf{Problem} & \textbf{Universe} & \textbf{Sense} & \textbf{Weights} & \textbf{Argument}\\
        \hline
        \multicolumn{5}{l}{\emph{Type A: the feasibility problem is $\NPS$-complete}} \\
        $\NPS$-complete & universe $\U$ & any & $\{0\}$ & \Cref{lem:NP-SandLOP-completeness} \\
        \textsc{Satisfiability} & literals $L$ & any & $\{1\}$ & \Cref{thm:cook-levin-clop} \\
        \textsc{Satisfiability} & variables $X$ & any & $\{1\}$ & \ref{compendium:satisfiability:variables} \\
        \textsc{3-Satisfiability} & literals $L$ & any & $\{1\}$ & \Cref{cor:constant-cardinality} \\
        \textsc{3-Satisfiability} & variables $X$ & any & $\{1\}$ & \ref{compendium:satisfiability:variables} \\
        \textsc{Directed Hamiltonian Path} & arcs $A$ & any & $\{1\}$ & \Cref{cor:constant-cardinality} \\
        \textsc{Undirected Hamiltonian Path} & edges $E$ & any & $\{1\}$ & \Cref{cor:constant-cardinality} \\
        \textsc{Directed Hamiltonian Cycle} & arcs $A$ & any & $\{1\}$ & \Cref{cor:constant-cardinality} \\
        \textsc{Undirected Hamiltonian Cycle} & edges $E$ & any & $\{1\}$ & \Cref{cor:constant-cardinality} \\
        \textsc{TSP (on general graphs)} & edges $E$ & any & $\{1\}$ & \Cref{cor:constant-cardinality} \\
        \textsc{Subset Sum} & elements $A$ & any & $\{1\}$ & \ref{compendium:subsetsum} \\
        \textsc{Partition} & elements $A$ & any & $\{1\}$ & \ref{compendium:partition} \\
        \textsc{$2$-Machine Makespan Scheduling} & job times $P$ & any & $\{1\}$ & \ref{compendium:scheduling} \\
        \textsc{$2$-Vertex-Disj. Directed Paths} & arcs $A$ & any & $\{1\}$ & \ref{compendium:2vdp} \\
        \textsc{$k$-Vertex-Disj. Directed Paths} & arcs $A$ & any & $\{1\}$ &  \ref{compendium:2vdp} \\
        \textsc{Maximum Benefit Set Cover} & sets $\mathcal{F}$ & any & $\{0\}$ & \ref{compendium:mbsc} \\
        \hline
        \multicolumn{5}{l}{\emph{Type B: feasibility can be decided in polynomial time}} \\
        \textsc{Vertex Cover} & vertices $V$ & min & $\{1\}$ & \Cref{prop:vc-lop-complete} \\
        \textsc{Dominating Set} & vertices $V$ & min & $\{1\}$ & \ref{compendium:ds} \\
        \textsc{Set Cover} & sets $\mathcal{F}$ & min & $\{1\}$ & \ref{compendium:sc} \\
        \textsc{Hitting Set} & elements $U$ & min & $\{1\}$ & \ref{compendium:sc} \\
        \textsc{Directed Feedback Vertex Set} & vertices $V$ & min & $\{1\}$ & \ref{compendium:fvs} \\
        \textsc{Feedback Arc Set} & arcs $A$ & min & $\{1\}$ & \ref{compendium:fas} \\
        \textsc{p-Center} & facilities $F$ & min & $\{1\}$ &  \ref{compendium:p-center-p-median} \\
        \textsc{p-Median} & facilities $F$ & min & $\{1\}$ &  \ref{compendium:p-center-p-median} \\
        \textsc{3-Dimensional Matching} & triples $T$ & max & $\{1\}$ & \ref{compendium:3dm} \\
        \textsc{Independent Set} & vertices $V$ & max & $\{1\}$ & \Cref{prop:vc-lop-complete} (duality) \\
        \textsc{Clique} & vertices $V$ & max & $\{1\}$ & \ref{compendium:clique:vertices} \\
        \textsc{Clique} & edges $E$ & max & $\{1\}$ & \ref{compendium:clique:edges} \\
        \textsc{Steiner Tree} & edges $E$ & min & $\{1\}$ & \ref{compendium:steinertree} \\
        \textsc{Knapsack} & items & max & $\Z_{\geq 0}$ & \ref{compendium:knapsack} \\
        \textsc{TSP (on complete graphs)} & edges $E$ & min & $\{1,2\}$ & \ref{compendium:tsp:complete} \\
        \textsc{TSP (on complete graphs)} & edges $E$ & min & $\{0,1\}$ & \ref{compendium:tsp:complete} \\
        \hline
    \end{tabular}
\caption{$\LOP$-complete problems. The column ``Weights'' gives the smallest weight class for which we establish $\LOP$-completeness; by \Cref{lem:weight-monotonicity}, completeness extends to every larger class, in particular to general weights $\Z_{\geq 0}$. For Type A problems, the sense is irrelevant. For Type B problems, the zero-weight version is polynomial-time decidable, and duality yields the corresponding problem with the opposite sense.}
\label{tab:LOP-complete-compendium}
\end{table}

\subparagraph*{Type A.}
For each Type A entry, the feasibility problem is $\NPS$-complete \cite{DBLP:journals/corr/abs-2602-12350}, so the zero-weight version is $\LOP$-complete by \Cref{lem:NP-SandLOP-completeness}.
For \textsc{3-Satisfiability} with literal universe and for the Hamiltonian path and cycle problems, including Traveling Salesman on general graphs, every solution has the same cardinality ($n$, $|V|-1$, and $|V|$, respectively), so the unit-weight version is complete by \Cref{cor:constant-cardinality}.
For the unit-weight versions of \textsc{Satisfiability (Variable Universe)}, \textsc{Subset Sum}, \textsc{Partition}, \textsc{$2$-Machine Makespan Scheduling}, \textsc{$2$-Vertex-Disj. Directed Paths}, and \textsc{$k$-Vertex-Disj. Directed Paths}, the solutions in general do not have constant cardinality.
For these, we give SE reductions from an $\NPS$-complete problem whose target instances have solutions of constant cardinality~$k$. Then $k$ is both a lower and an upper bound on the weight of every feasible solution, and \Cref{lem:bound} yields completeness of
the unit-weight version for both senses.

\subparagraph*{Type B.}
Entries (B1)--(B6) and (B9)--(B11) follow from \Cref{lem:bound}, applied to an SE reduction from an $\NPS$-complete problem to the derived $\NPS$ problem.
For each, we state the reduction and the bound.
Entries (B7) and (B8) follow from tight solution-embedding reductions from other entries together with \Cref{lem:TSE-transitive}.
The reductions in (B1)--(B5) start from the graph $G$ of \Cref{prop:vc-lop-complete}, in which every vertex cover has at least $k := n + 2m$ vertices and the covers of size exactly $k$ correspond to the satisfying assignments.
Composing with that reduction gives an SE reduction from \textsc{3-Satisfiability}.

\begin{description}
    \comitem{(A1)}{compendium:satisfiability:variables} \textsc{Satisfiability} and \textsc{3-Satisfiability}.
    We describe a self-reduction for \textsc{Satisfiability} (the same holds for \textsc{3-Satisfiability}).
    Let $L$ be the set of literals of the original instance.
    We introduce a variable $x_\ell$ for every literal $\ell \in L$ and replace every occurrence of $\ell$ in a clause by the positive literal $x_\ell$.
    For every pair of complementary literals $\ell, \overline{\ell}$, we add the clauses
    $$
        (x_\ell \lor x_{\overline\ell}) \land (\overline{x}_\ell \lor \overline{x}_{\overline\ell}),
    $$
    which express $x_\ell \oplus x_{\overline\ell}$, i.e., exactly one of the two variables is true.
    The embedding $f(\ell) = x_\ell$ maps the satisfying assignments $L_\alpha$ of the original instance bijectively onto the sets of true variables of the new instance.
    Since every satisfying assignment contains exactly one literal of each of the $|L|/2$ complementary pairs, every solution of the new instance contains exactly $|L|/2$ variables, so \Cref{lem:bound} applies.
    
    \comitem{(A2)}{compendium:subsetsum} \textsc{Subset Sum}.
    We modify the reduction from \textsc{3-Satisfiability} to \textsc{Subset Sum} by Sipser~\cite{DBLP:books/daglib/0086373}.
    Given a \textsc{3-Satisfiability} instance $I=(L,C)$, we construct a \textsc{Subset Sum} instance $(S,M)$ using a positional number system.
    For each literal $\ell \in L$, we introduce a number $a_\ell$ with a digit 1 in the column corresponding to its variable and additional digits 1 in the columns of the clauses containing $\ell$.
    For each clause $c \in C$, we introduce three auxiliary numbers with respective digits 11, 12, and 13 in the column corresponding to c.
    The target $M$ has digit 1 in every variable column and digit 14 in every clause column.
    We choose a sufficiently large base to prevent carries between columns.
    
    The variable columns force the selection of exactly one number from each complementary literal pair.
    In each clause column, the selected literal numbers contribute $k \in \{0,1,2,3\}$.
    If $k=0$, the target cannot be reached; if $k \in \{1,2,3\}$, exactly one auxiliary number, with digit $14-k$, completes the target.
    Since the auxiliary digits are 11, 12, and 13, no two auxiliary numbers can be selected for the same clause.
    Hence, a subset sums to $M$ if and only if the selected literals form a satisfying assignment.
    
    Moreover, every feasible subset contains exactly one number per variable and exactly one auxiliary number per clause. Consequently, every Subset Sum solution has cardinality
    $$
        |L|/2+|C|,
    $$
    establishing the desired constant cardinality property, so \Cref{lem:bound} applies.
    
    \comitem{(A3)}{compendium:partition} \textsc{Partition}.
    We use the reduction from \textsc{Subset Sum} to \textsc{Partition} by Karp~\cite{DBLP:conf/coco/Karp72}.
    Let $I = (\{a_1,\ldots,a_n\},M)$ be a \textsc{Subset Sum} instance, and let $A=\sum^n_{i=1} a_i$.
    Without loss of generality, we require the last element of a \textsc{Partition} solution to be selected and thus in the first set of the partition, thereby breaking the symmetry between complementary solutions.
    We construct a \textsc{Partition} instance by retaining the original numbers and adding $M+1$ and $A+1-M$, where the latter is designated as the last element $a_n$.
    The embedding $f_I(a_i) = a_i$ for the original elements establishes the solution-embedding function:
    a subset of the original numbers sums to $M$ if and only if the corresponding \textsc{Partition} instance admits a partition in which the designated element belongs to the appropriate side.

    The solution-embedding $f_I$ also gives us constant cardinality:
    a \textsc{Partition} solution consists of the \textsc{Subset Sum} solution together with the last element $A+1-M$.
    Hence \Cref{lem:bound} applies.
    
    \comitem{(A4)}{compendium:scheduling} \textsc{Two-Machine Makespan Scheduling}.
    Here, the objective is a cost on the jobs assigned to the first machine, and the processing times are certain.
    Note that this differs from the variant with uncertain processing times and the makespan as objective.
    Given a \textsc{Partition} instance  $I = \{a_1,\ldots,a_n\}$, we construct a \textsc{Two-Machine Makespan Scheduling} instance by introducing one job with processing time $p_i=a_i$ for each $a_i$, and setting the makespan threshold to $T = \frac{1}{2} \sum_{i=1}^n a_i$.
    A feasible schedule assigns the jobs to two machines such that the load on each machine is at most $T$.
    Since the total processing time is $2T$, both machines must have load exactly $T$, establishing a one-to-one correspondence with feasible partitions.
    We define the solution-embedding by $f_I(a_i) = p_i$ and, without loss of generality, require the last element to be assigned to the first machine to break the symmetry between the two machines.

    Moreover, the reduction preserves cardinality:
    the number of jobs assigned to the first machine equals the number of elements in the corresponding designated partition subset.
    Since the preceding reductions guarantee a fixed cardinality for feasible solutions, this property carries over to the scheduling instance.
    Thus, \Cref{lem:bound} applies.
    
    \comitem{(A5)}{compendium:2vdp} \textsc{Two Vertex Disjoint Directed Path} and \textsc{$k$-Vertex Disjoint Directed Path}.
    We use the reduction from \textsc{3-Satisfiability} to \textsc{Two Vertex Disjoint Directed Path} by Fortune, Hopcroft, and Wyllie~\cite{DBLP:journals/tcs/FortuneHW80}.
    A summary of this reduction can be found in \cite{DBLP:conf/ipco/GruneW25}.
    This reduction can be constructed such that every feasible traversal of each switch, lobe, and clause gadget uses a fixed number of arcs, independently of the choices made within the gadget.
    Consequently, the total number of arcs used by the two paths depends only on the constructed instance and not on the particular satisfying assignment.
    Therefore, \Cref{lem:bound} applies.
    The same also applies for the generalization to \textsc{$k$-Vertex Disjoint Directed Path}.

    \comitem{(A6)}{compendium:mbsc} \textsc{Maximum Benefit Set Cover}.
    Note that \textsc{Maximum Benefit Set Cover} is the $\LOP$ problem of the derived $\NPS$ problem of \textsc{Set Cover}, since only an additional linear objective (the benefit) is introduced.
    Since \textsc{Set Cover} is $\LOP$-complete, the derived $\NPS$ problem of \textsc{Set Cover} is $\NPS$-complete by \Cref{lem:LOP-complete-derived-NPS}.
    Applying \Cref{lem:NP-SandLOP-completeness} yields $\LOP$-completeness of \textsc{Maximum Benefit Set Cover} for both $\sigma \in \{\min,\max\}$ and $b \equiv 0 \equiv t$, where $b$ is the benefit weight function.

    \comitem{(B1)}{compendium:ds} \textsc{Dominating Set.}
    Apply the construction of \Cref{sec:LOPcompleteness} to $G$, which has no isolated vertices.
    By the second statement shown there, every dominating set $D$ yields a vertex cover of size at most $|D|$, with strict inequality if $D$ contains a gadget vertex.
    Hence every dominating set has at least $k$ vertices, and the dominating sets of size $k$ are exactly the vertex covers of size $k$.

    \comitem{(B2)}{compendium:sc} \textsc{Set Cover} and \textsc{Hitting Set}.
    We use the reductions by Karp \cite{DBLP:conf/coco/Karp72}.
    For \textsc{Set Cover}, take the edges of $G$ as ground set and, for each vertex $v$, the set of edges incident to $v$; the universe consists of the indices of these sets.
    For \textsc{Hitting Set}, take the vertices of $G$ as elements and its edges as the sets to be hit.
    In both cases, the solutions correspond exactly to the vertex covers of $G$, so the bound $k$ carries over.

    \comitem{(B3)}{compendium:fvs} \textsc{Directed Feedback Vertex Set.}
    The reduction by Karp \cite{DBLP:conf/coco/Karp72} works as follows.
    Replace each edge $\{u,v\}$ of $G$ by the two arcs $(u,v)$ and $(v,u)$, and call the resulting digraph $D$. Every arc of $D$ lies on a $2$-cycle, so a feedback vertex set must be a vertex cover of $G$; conversely, removing a vertex cover leaves no arcs and hence no cycles. Thus the feedback vertex sets of $D$ are exactly the vertex covers of $G$, and the bound $k$ carries over.

    \comitem{(B4)}{compendium:fas} \textsc{Feedback Arc Set.}
    We again use a reduction by Karp \cite{DBLP:conf/coco/Karp72}.
    Start from the digraph $D$ of \ref{compendium:fvs}.
    Split each vertex $v$ into an arc $a_v = (v^-, v^+)$, and replace each arc $(u,v)$ of $D$ by two internally disjoint paths of length two from $u^+$ to $v^-$. All arcs have weight $1$, and $f$ maps a literal $\ell$ to $a_\ell$.
    Let $F$ be any feedback arc set. Let $X$ consist of the vertices $v$ with $a_v \in F$ and the heads $v$ of all connections $(u,v)$ whose two paths both contain an arc of $F$.
    Then $X$ is a feedback vertex set of $D$: a cycle of $D$ avoiding $X$ would lift to a cycle avoiding $F$, by using for each connection a path without an arc of $F$.
    Each vertex of $X$ is charged either to its split arc or to two non-split arcs of $F$, so $|X| \leq s + r/2$, where $s$ and $r$ are the numbers of split and non-split arcs in $F$.
    Hence $|F| = s + r \geq |X| \geq k$, which is the bound. If $|F| = k$, then $r = 0$, so $F = \{a_v : v \in X\}$ for a feedback vertex set $X$ of size $k$. Conversely, $\{a_v : v \in X\}$ is a feedback arc set for every feedback vertex set $X$. Thus the solutions of weight $k$ correspond exactly to the vertex covers of $G$ of size $k$.

    \comitem{(B5)}{compendium:p-center-p-median} \textsc{p-Center} and \textsc{p-Median}.
    Here we consider the variants in which the number of opened facilities is the objective and every customer must be served within a given radius.
    Note that this differs from the variant with fixed $p$ and the radius as objective.
    We consider a modified version of the reduction by Cornuéjols, Nemhauser, and Wolsey \cite{cornuejols1983uncapicitated} from \textsc{Vertex Cover} to \textsc{p-Center}.
    Given a \textsc{Vertex Cover} instance $I=((V,E),k)$, the reduction constructs a \textsc{p-Center} instance $I'=(F,C,c,p,k')$.
    We identify each vertex $v \in V$ with a facility $v \in F$ and each edge $e \in E$ with a customer $e \in C$.
    The service costs are defined as $c(v,e) = 0$ if $v \in e$ and $c(v,e)=|V|+1$ otherwise.
    Finally, we set $p=k$ and $k'=0$.
    
    The reduction is solution-embedding, mapping each vertex $v \in V$ to the corresponding facility $f_I(v) = v \in F$.
    Since the constraint $k'=0$ requires every customer to be served at cost zero, the selected facilities must correspond to a vertex cover of $G$.

    The solution size is defined as the number of selected facilities, and the objective is to minimize this number subject to the constraint $\max_{e\in C} \min_{v\in F} c(v, e) \leq k'$.
    Since $|F'|$ is the objective function, and $p = k$ is the threshold, this problem has unit weights.
    Given the one-to-one correspondence between vertex covers and facility sets, $p$ is tight if $k$ is tight.

    The reduction for \textsc{p-Median} is the same.
    The only difference between \textsc{p-Center} and \textsc{p-Median} is the customer constraint.
    We substitute $\max_{e\in C} \min_{v\in F} c(v, e) \leq k'$ by $\sum_{e \in C} \min_{v \in F} c(v,e) \leq k'$.
    With the same reduction, $k'$ still needs to be $0$, which does not change the underlying logic.

    \comitem{(B6)}{compendium:3dm} \textsc{3-Dimensional Matching.}
    Here the feasible solutions are all matchings, and the sense is $\max$.
    With ground sets of size $q$, every matching has at most $q$ triples, so $t = q$ is an upper bound on the weight of every feasible solution, and the threshold-feasible solutions are the perfect matchings.
    The SE reduction from \textsc{Satisfiability} to perfect matchings in \cite{DBLP:journals/corr/abs-2602-12350} therefore satisfies \Cref{lem:bound}.

    \comitem{(B7)}{compendium:clique:vertices} \textsc{Clique (Vertex-based)}.
    A vertex set is a clique of a graph if and only if it is an independent set of its complement.
    Complementing the graph, with the identity on the vertices, maps the solutions of \textsc{Independent Set} bijectively onto those of \textsc{Clique} and preserves unit weights and thresholds.
    Hence it is a tight solution-embedding reduction from \textsc{Maximum Cardinality Independent Set}, which is $\LOP$-complete by \Cref{prop:vc-lop-complete}, and \textsc{Clique} is $\LOP$-complete by \Cref{lem:TSE-transitive}.
    
    \comitem{(B8)}{compendium:clique:edges} \textsc{Clique (Edge-based)}.
    We consider a reduction from \textsc{Clique (Vertex-based)} to \textsc{Clique (Edge-based)} in \cite{DBLP:journals/corr/abs-2602-12350}.
    Given an instance $I=(G,k)$, where $G=(V,E)$, the reduction constructs an instance $I'=(G',k')$, where $G'=(V \cup V', E')$ and $k'=\binom{2k}{2}$.
    Here, $V'$ is a copy of $V$.
    For every vertex $v \in V$, the construction introduces an edge $\{v,v'\} \in E'$.
    Furthermore, for every edge $\{v,w\} \in E$, the construction adds all four edges $\{v,w\},\{v',w\},\{v,w'\}$, and $\{v',w'\}$ to $E'$.

    There is a one-to-one correspondence between cliques of size $k$ in $G$ and cliques of size $2k$ in $G'$.
    The reduction is solution-embedding, with the embedding function mapping each vertex $v \in V$ to the edge $f_I(v) = \{v,v'\} \in E'$.

    By construction, every edge in the target instance has unit weight.
    Hence, the weight of any (edge-based) clique equals its number of edges.
    Given that $k$ is a tight threshold, every threshold-feasible solution in $G'$ is a clique of exactly $2k$ vertices and therefore has $\binom{2k}{2}$ edges.
    Thus, $k'$ is tight as well and all threshold-feasible solutions have the same size.

    \comitem{(B9)}{compendium:steinertree} \textsc{Steiner Tree}.
    We consider the folklore reduction from \textsc{3-Satisfiability} to \textsc{Steiner Tree} described in \cite{DBLP:journals/corr/abs-2602-12350}.
    Let $I=(L,C)$ be a \textsc{3-Satisfiability} instance, where $L$ denotes the set of literals and $C$ the set of clauses.
    The reduction constructs a \textsc{Steiner Tree} instance with designated terminals $s$ and $t$, a Steiner vertex for every literal $\ell \in L$, and additional Steiner vertices connecting consecutive literal pairs in a diamond chain.
    Specifically, defining $v_0 := s$, each pair of complementary literals is connected between consecutive vertices $v_{i-1}$ and $v_i$, with the final pair connected to $t$.

    For every clause $c_j \in C$, the construction introduces a terminal vertex $c_j$, which is connected to each literal occurring in the clause by a path of length $|L|+1$.
    All edges have weight one, and the target budget is $k = |L|+|C|(|L|+1)$.
    
    The construction is correct because any Steiner tree of weight at most $k$ must connect $s$ to $t$ through the diamond chain:
    a path through a clause terminal would be too long.
    To see this, note that every clause terminal $c_j$ is joined to the rest of the graph only by its clause paths, whose inner vertices have degree two.
    Hence every Steiner tree contains at least one complete clause path for each $c_j$, i.e., at least $|C|(|L|+1)$ edges, and these paths are pairwise edge-disjoint and disjoint from the diamond chain.
    If the $s$-$t$ path of the tree does not run entirely through the diamond chain, it leaves the chain at a literal vertex and returns via some clause terminal, and thus uses a second complete clause path of that terminal.
    This costs at least $|L|+1$ further edges, so the tree has more than $|L| + |C|(|L|+1) = k$ edges.
    Otherwise, the $s$-$t$ path uses exactly two edges per diamond, i.e., $|L|$ edges.
    In both cases, every Steiner tree has at least $k$ edges, and a tree with exactly $k$ edges selects one literal per variable and attaches each clause terminal by a single path to a selected literal.
    The selected literal edges encode a truth assignment, with a positive literal selected if and only if its corresponding edge $\{v_{i-1},\ell_i\}$ belongs to the tree, and analogously for negative literals.
    The budget therefore ensures that every clause is satisfied.
    Conversely, every satisfying assignment yields a Steiner tree of weight at most $k$.
    
    The reduction is solution-embedding: map each literal $\ell$ to its corresponding edge $f_I(\ell) = \{v_{i-1},\ell\}$.

    By construction, every edge in the target instance has weight one, so the weight of any Steiner tree equals its number of edges.
    By the arguments from above, every feasible solution contains at least $|L|$ edges connecting $s$ to $t$ through the diamond chain and at least $|C|(|L|+1)$ additional edges to connect the clause terminals.
    Thus, $k$ is a lower bound on the weight of every feasible solution, and \Cref{lem:bound} applies.
    
    \comitem{(B10)}{compendium:knapsack} \textsc{Knapsack.}
    Reduce from \textsc{Subset Sum} with numbers $a_i$ and target $B$.
    Take items with sizes $s_i = a_i$, capacity $B$, values $w_i = a_i$, threshold $t = B$, and the identity on the items.
    Every feasible solution satisfies $w(S) = s(S) \leq B$, so $t$ is an upper bound, and the threshold-feasible solutions are exactly the subsets with sum $B$.
    Hence \Cref{lem:bound} applies.
    Since the values are part of the weights, the entry is for general weights.
    The unit-weight version is solvable in polynomial time by choosing the smallest items, so it is not $\LOP$-complete unless $\PTIME = \NP$.

    \comitem{(B11)}{compendium:tsp:complete} \textsc{Traveling Salesman.}
    We consider the version restricted to complete graphs with edge weights in $\{1,2\}$, which is metric.
    Reduce from \textsc{Undirected Hamiltonian Cycle} on $G = (V,E)$ with $|V| \geq 3$.
    Take the complete graph on $V$ with weight $1$ on the edges of $E$ and weight $2$ on all other edges, threshold $t = |V|$, and the identity on $E$.
    Every tour has $|V|$ edges of weight at least $1$, so $t$ is a lower bound, and \Cref{lem:bound} applies.
    The tours of weight $|V|$ are exactly the Hamiltonian cycles of $G$.
    The same argument works with weights in $\{0,1\}$ (weight $0$ on $E$, threshold $0$), but these weights are not metric.
    With unit weights, all tours have the same weight, so the unit-weight version is not $\LOP$-complete unless $\PTIME = \NP$.
    
\end{description}

\subparagraph*{Problems not included.}
For \textsc{Uncapacitated Facility Location}, the natural objective depends on the distances from the customers to the chosen facilities and is therefore not linear in the universe of facilities.
This problem does not fit the $\LOP$ framework in its natural form, and we omit it.

\newpage

\bibliography{src/bib/general,src/bib/nps,src/bib/regret,src/bib/reductions}

\end{document}